%% file: KK3d_ver35bib.tex
\documentclass[ALICE,manyauthors]{cernphprep}

\usepackage[pdftex,pdfborderstyle={0 0 1}]{hyperref}
\hypersetup{pdfborder=0 0 1}

\usepackage{lineno}
\usepackage{multirow} 
\usepackage{ltxcmds}
\usepackage{placeins}

\usepackage[nospace]{cite}

\newcommand{\bibname}{References}
\usepackage[comma,square,numbers,sort&compress]{natbib}
\usepackage{color}
\usepackage{hyperref}
\hypersetup{
    colorlinks=false,
    pdfborder=kTRUE,
    linkcolor=blue,
    filecolor=magenta,
    urlcolor=cyan,
}

\newcommand{\Kzs}{\rm K^{ 0}_S}
\newcommand{\Kpm}{\rm K^{\pm}}

\newcommand{\Kaon}{\rm K}

\newcommand{\pro}{\rm p}
\newcommand{\pbar}{\rm \overline{p}}
\newcommand{\kzs}{\rm K^0_S}
\newcommand{\kz}{\rm K^0}
\newcommand{\kzb}{\rm \overline{K}{}^0}

\newcommand{\pvec}[1]{\vec{#1}\mkern2mu\vphantom{#1}}
\newcommand{\kstar}{k^*}
\newcommand{\vkstar}{\pvec{k}^*}
\newcommand{\rstar}{r^*}
\newcommand{\vrstar}{\pvec{r}^*}

\begin{document}%

\begin{titlepage}
\PHyear{2017}
\PHnumber{185}      
\PHdate{21 July}  
%

\title{Kaon femtoscopy in Pb--Pb collisions at $\sqrt{s_{\rm{NN}}}$ = 2.76 TeV}
\ShortTitle{Kaon femtoscopy  in Pb--Pb at $\sqrt{s_{\rm{NN}}}$ = 2.76 TeV}   

\Collaboration{ALICE Collaboration\thanks{See Appendix~\ref{app:collab} for the list of collaboration members}}
\ShortAuthor{ALICE Collaboration} 

\begin{abstract}
We present the results of three-dimensional femtoscopic analyses for charged and neutral kaons recorded by ALICE
in Pb--Pb collisions at $\sqrt{s_{\rm{NN}}}$ = 2.76 TeV.
Femtoscopy is used to measure the space-time characteristics of particle production from the effects of quantum statistics and final-state interactions in two-particle correlations.
Kaon femtoscopy is an important supplement to
that of pions because it allows one to distinguish between different model scenarios working
equally well for pions.
In particular, we compare the measured 3D kaon radii with a purely hydrodynamical calculation and a model where the hydrodynamic phase is followed by a hadronic rescattering stage.
The former predicts an approximate transverse mass ($m_{\mathrm{T}}$) scaling of source radii obtained from
pion and kaon correlations. This $m_{\mathrm{T}}$ scaling appears to be broken in our data, which
indicates the importance of the hadronic rescattering phase at LHC energies.
A $k_{\mathrm{T}}$ scaling of pion and kaon source radii is observed instead.
The time of maximal emission of the system is estimated using the three-dimensional femtoscopic
analysis for kaons. The measured emission time is larger than that of pions. Our observation is
well supported by the hydrokinetic model predictions.

\end{abstract}
\end{titlepage}
\setcounter{page}{2}
\section{Introduction}

Extremely high energy densities achieved in heavy-ion collisions at the Large Hadron
Collider (LHC) are expected to lead to the formation of the quark-gluon plasma (QGP), a state characterized by partonic degrees of freedom \cite{Cabibbo:1975ig,Shuryak:1980tp}. The systematic study of many observables (transverse momentum spectra, elliptic flow, jets, femtoscopy correlations) measured at the Relativistic Heavy Ion Collider (RHIC)
 and the LHC confirmed the presence of strong collective motion and the hydrodynamic behavior of the system (see e.g. \cite{RHIC_hydro_a,RHIC_hydro_b,RHIC_hydro_c,RHIC_hydro_d}
and \cite{ALICE_hydro_a,ALICE_hydro_b,ALICE_hydro_c,ALICE_hydro_d}). Whereas since quite a long time hydrodynamics describes momentum based observables, it could not describe spatial distributions at decoupling.
Correlation femtoscopy (commonly referred to as \emph{femtoscopy} or \emph{HBT, Hanbury Brown and Twiss interferometry}), measures the space-time characteristics of particle production using particle correlations due to the effects of quantum statistics and strong and Coulomb final-state interactions~\cite{Goldhaber:1960sf,Kopylov:1972qw,Kopylov:1974uc,Lednicky:2005af,Lisa:2005dd}.
The problem to describe the spatio-temporal scales derived from femtoscopy in heavy-ion collisions at RHIC was solved only a few years ago, strongly constraining the hydrodynamical models \cite{Sinyukov:2009ce,Broniowski:2008vp,Vredevoogd:2009zu}.
 The following factors were understood to be important: existence of prethermal transverse flow, a crossover transition between quark-gluon and hadron matter, non-hydrodynamic behavior of the hadron gas at the latest stage (hadronic cascade phase), and correct matching between hydrodynamic and non-hydrodynamics phases (see e.g. \cite{Sinyukov:2009ce}).

New challenges for hydrodynamics appeared when data were obtained at the LHC: the large statistics now allows one to investigate not only pion femtoscopy, which is the most common femtoscopic analysis, but also femtoscopy of heavier particles in differential analyses with high precision.

The main objective of ALICE \cite{Aamodt:2008zz} at the LHC is to study the QGP.
ALICE has excellent capabilities to study femtoscopy observables due to good track-by-track particle identification (PID), particle acceptance down to low transverse momenta $p_{\rm T}$, and good resolution of secondary vertices.
We already studied pion correlation radii in Pb--Pb collisions at
2.76~TeV \cite{Aamodt:2011mr,Adam:2015vna}.
Pion femtoscopy showed genuine effects originating from collective flow
 in heavy-ion collisions, manifesting as a decrease of the source radii with increasing pair transverse mass $m_{\rm T} = \sqrt{k_{\rm T}^2+m^2}$~\cite{Lisa:2005dd,Bearden:2000ex}, where $k_{\rm T}=|\mathbf{p}_{\mathrm{T,1}}+\mathbf{p}_{\mathrm{T,2}}|/2$ is the average  transverse momentum of the corresponding pair and $m$ is the particles mass.

The next most numerous particle species after pions are kaons.
The kaon analyses are expected to offer a cleaner signal compared to pions, as they are less affected by resonance decays.
Studying charged and neutral kaon correlations together provides a convenient experimental consistency check, since they require different detection techniques.
The theoretical models which describe pion femtoscopy well, should
describe kaon results with equal precision.

Of particular interest is the study of the $m_{\rm T}$-dependence of pion and kaon source radii. It was shown that the hydrodynamic picture of nuclear collisions for the particular case of small transverse flow leads to the same $m_{\rm T}$ behavior of the longitudinal radii ($R_{\rm long}$) for pions and kaons~\cite{Makhlin:1987gm}. This common $m_{\rm T}$-scaling for $\pi$ and $\Kaon$ is an indication that the thermal freeze-out occurs simultaneously for $\pi$ and $\Kaon$ and that these two particle species are subject to the same velocity boost from collective flow.
Previous kaon femtoscopy studies carried out in Pb--Pb collisions at the SPS by the NA44 and NA49 Collaborations~\cite{Bearden:2001sy, Afanasiev:2002fv} reported the decrease of $R_{\rm long}$ with $m_{\rm T}$ as $\sim m_{\rm T}^{-0.5}$ as a consequence of the boost-invariant longitudinal flow. Subsequent studies carried out in Au--Au collisions at RHIC~\cite{Adams:2003xp,Adamczyk:2013wqm,Abelev:2006gu,Afanasiev:2009ii} have shown the same power in the $m_{\rm T}$-dependencies for $\pi$ and $\Kaon$ radii, consistent with a common freeze-out hypersurface. Like in the SPS data, no exact universal $m_{\rm T}$-scaling for the 3D radii was observed at RHIC, but still these
experiments observed an approximate $m_{\rm T}$-scaling for pions and kaons.
The recent study of the $m_{\rm T}$-dependence of kaon three-dimensional radii
performed by the PHENIX collaboration \cite{Adare:2015bcj} demonstrated breaking of this scaling especially for the ``long'' direction.
PHENIX reported that the Hydro-Kinetic Model (HKM) describes well the overall trend of femtoscopic radii for pions and kaons \cite{Karpenko:2012yf,Karpenko:2010te}.

We have published previously the study of one-dimensional
correlation radii of different particle species:  $\pi^{\pm}\pi^{\pm}$, $\Kpm\Kpm$, $\Kzs\Kzs$, $\pro\pro$, and $\pbar\pbar$ correlations in Pb--Pb collisions at $\sqrt{s_{\mathrm {NN}}}=2.76$~TeV for several intervals of centrality and transverse mass \cite{Adam:2015vja}. The decrease of the source radii with increasing transverse mass was observed for all types of particles, manifesting a fingerprint of collective flow in heavy-ion collisions. The one-dimensional femtoscopic radii demonstrated the approximate $m_{\rm T}$-scaling as it was expected by hydrodynamic model considerations~\cite{Lisa:2005dd}.

Recent calculations made within a 3+1D hydrodynamical model coupled with a statistical hadronization code taking into account the resonance contribution, THERMINATOR-2, showed the approximate scaling of the three dimensional radii with transverse mass for pions, kaons and protons \cite{Kisiel:2014upa}.
An alternative calculation, that is the Hydro-Kinetic Model, including a hydrodynamic phase as well as a hadronic rescattering stage, predicts violation of such a scaling between pions and kaons at LHC energies \cite{Shapoval:2014wya}.
Both models observe approximate scaling if there is no rescattering phase.
It is suggested in \cite{Shapoval:2014wya} that rescattering has significantly
different influence on pions and kaons and is responsible for the violation of
$m_{T}$-scaling at the LHC energies.
Moreover, the analysis of the emission times of pions and kaons obtained within HKM in \cite{Sinyukov:2015kga} showed that kaons are emitted later than pions due to rescattering through the rather long-lived ${\rm K}^{*}(892)$ resonance.
This effect can explain the  $m_{\rm T}$-scaling violation predicted in \cite{Shapoval:2014wya}.

In \cite{Shapoval:2014wya} it was found that immediately decaying the ${\rm K}^{*}(892)$ and $\phi$(1020) resonances at the chemical freeze-out hypersurface has only a negligible influence on the kaon radii. In this scenario, resonances were allowed to be regenerated in the hadronic phase. Further analysis in \cite{Sinyukov:2015kga} showed that it is indeed the regeneration of the  ${\rm K}^{*}(892)$ resonance through hadronic reactions which is responsible for the $m_{\rm T}$-scaling violation predicted in \cite{Shapoval:2014wya}. This mechanism clearly manifests itself in the prolonged emission time of kaons caused by the rather long lifetime of the ${\rm K}^{*}(892)$ resonance \cite{Shapoval:2014wya}.

The approximate scaling of pion and kaon radii was predicted by investigating 3+1D hydrodynamical model + THERMINATOR-2 in \cite{Kisiel:2014upa} to hold for each of the three-dimensional radii separately. The scaling of one-dimensional pion and kaon radii was also studied in \cite{Kisiel:2014upa}.
It was shown that after averaging the three-dimensional radii and taking into account a mass-dependent Lorentz-boost factor, a deviation between one-dimensional pion and kaon radii appeared.
These circumstances made it impossible to discriminate between THERMINATOR-2 \cite{Kisiel:2014upa} and the HKM calculation \cite{Shapoval:2014wya} in the earlier published one-dimensional analysis of pion and kaon radii by ALICE \cite{Adam:2015vja}. The three-dimensional study presented here is not impeded by these effects and allows one to discriminate between the hypothesis of approximate scaling of three-dimensional radii predicted in \cite{Kisiel:2014upa} and strong scaling violation proposed in \cite{Shapoval:2014wya}. Thus the study of the $m_{\rm T}$-dependence of three-dimensional pion and kaon radii can unambiguously distinguish between the different freeze-out scenarios and clarify the existence of a significant hadronic phase.

One more interesting feature of femtoscopy studies of heavy-ion collisions
concerns the ratio of radius components in the transverse plane.
The strong hydrodynamic flow produces significant positive space-time correlations during the evolution of the freeze-out hypersurface.
This influences the extracted radius parameters of the system in the plane perpendicular to the beam axis. The radius along the pair transverse momentum is reduced by the correlation with respect to the perpendicular one in the transverse plane. This effect appears to be stronger at LHC than at RHIC energies \cite{Kisiel:2008ws,Karpenko:2009wf}. It was studied by the ALICE collaboration for pions in Pb--Pb collisions at 2.76~TeV \cite{Adam:2015vna} at different centralities.
This work extends this study to kaons and compares the obtained transverse radii
with those found in the analysis for pions and to the model calculations discussed above.
The paper is organized as follows. Section 2 explains the data selection and describes the identification of charged and neutral kaons. In Section 3 the details of the analysis of the correlation functions are discussed together with the investigation of the systematic uncertainties. Section 4 presents the measured source radii as well as the extracted emission times and compares them to model predictions.
 Finally, Section 5 summarizes the obtained results and discusses them within the hydrokinetic approach.

\section{Data selection}
\label{sec:details}

Large sets of data were recorded by the ALICE collaboration at $\sqrt{s_{\rm NN}}=2.76$~TeV in Pb--Pb collisions. The about 8 million events from 2010 (used only in the $\Kzs\Kzs$ analysis) and about 40 million events from 2011 made it possible to perform the three-dimensional analyses of neutral and charged kaon correlations differentially in centrality and pair transverse momentum $k_{\rm T}$.
Three trigger types were used: minimum bias, semi-central (10-50\% collision centrality), and central (0-10\% collision centrality)~\cite{Abelev:2013qoq}.
The analyses were performed in the centrality ranges: (0--5\%), (0--10\%), (10--30\%), and (30--50\%).
The centrality was determined using the measured amplitudes in the V0 detector~\cite{Abelev:2013qoq}.
The following transverse momentum $k_{\mathrm{T}}$ bins were
considered: (0.2--0.4), (0.4--0.6), and (0.6--0.8) GeV/$c$ for charged kaons and
 (0.2--0.6), (0.6--0.8), (0.8--1.0), and (1.0--1.5) GeV/$c$ for neutral kaons.

Charged particle tracking is generally performed using the Time Projection Chamber (TPC)~\cite{Dellacasa:2000bm} and the Inner Tracking System (ITS)~\cite{Aamodt:2008zz}. The ITS also provides high spatial resolution in determining the primary collision vertex.

Particle identification (PID) for reconstructed tracks was carried out using both the TPC and the Time-of-Flight (TOF) detector~\cite{Cortese:545834}.
For TPC PID, a parametrization of the Bethe-Bloch formula was employed to calculate the specific energy loss (d$E$/d$x$) in the detector expected for a particle with a given mass and momentum. For PID with TOF, the particle mass hypothesis was used to calculate the expected time-of-flight as a function of track length and momentum.
For each PID method, a value $N_{\sigma}$ was assigned to each track denoting the number of standard deviations between the measured track d$E$/d$x$ or time-of-flight and the calculated one as described above.
Different cut values of $N_{\sigma}$ were chosen based on detector performance
 for various particle types and track momenta (see Table~\ref{tab:K0cuts} for
specific values used in both analyses). More details on PID can be found in
Secs. 7.2--7.5 of \cite{Abelev:2014ffa}.

The analysis details for charged and neutral kaons  are discussed separately below.
All major selection criteria are also  listed in Table~\ref{tab:K0cuts}.

\subsection{Charged kaon selection}

Track reconstruction for the charged kaon analysis was performed using the tracks' signal in the TPC. The TPC is divided by the central electrode into two halves, each of them is composed of 18 sectors (covering the full azimuthal angle) with 159 padrows placed radially in each sector. A track signal in the TPC consists of space points (clusters), each of which is reconstructed in one of the padrows.
A track was required to be composed out of at least 70 such clusters. The parameters of the track are determined by performing a Kalman fit to a set of clusters with an additional constraint that the track passes through the primary vertex. The quality of the fit is requested to have $\chi^2/{\mathrm{NDF}} $ better than 2. The transverse
momentum of each track was determined from its curvature in the uniform magnetic field.  The momentum from this fit in the TPC was used in the analysis.
Tracks were selected based on their distance of closest
approach (DCA) to the primary vertex, which was required to be less than 2.4~cm
in the transverse and less than 3.0~cm in the longitudinal direction.

\begin{figure}[t!]
  \begin{center}
    \includegraphics[width=.49\textwidth]{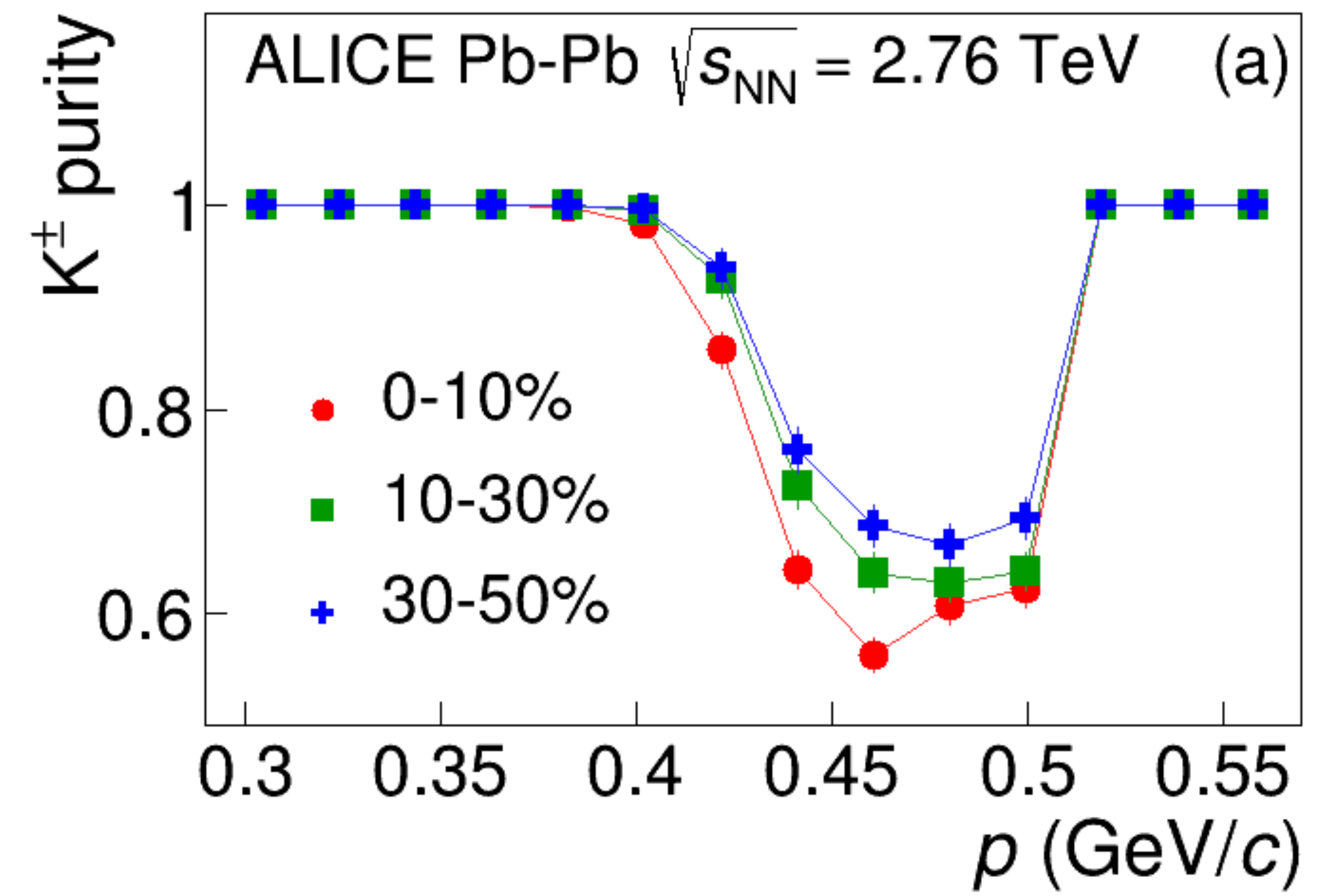}
    \includegraphics[width=.49\textwidth]{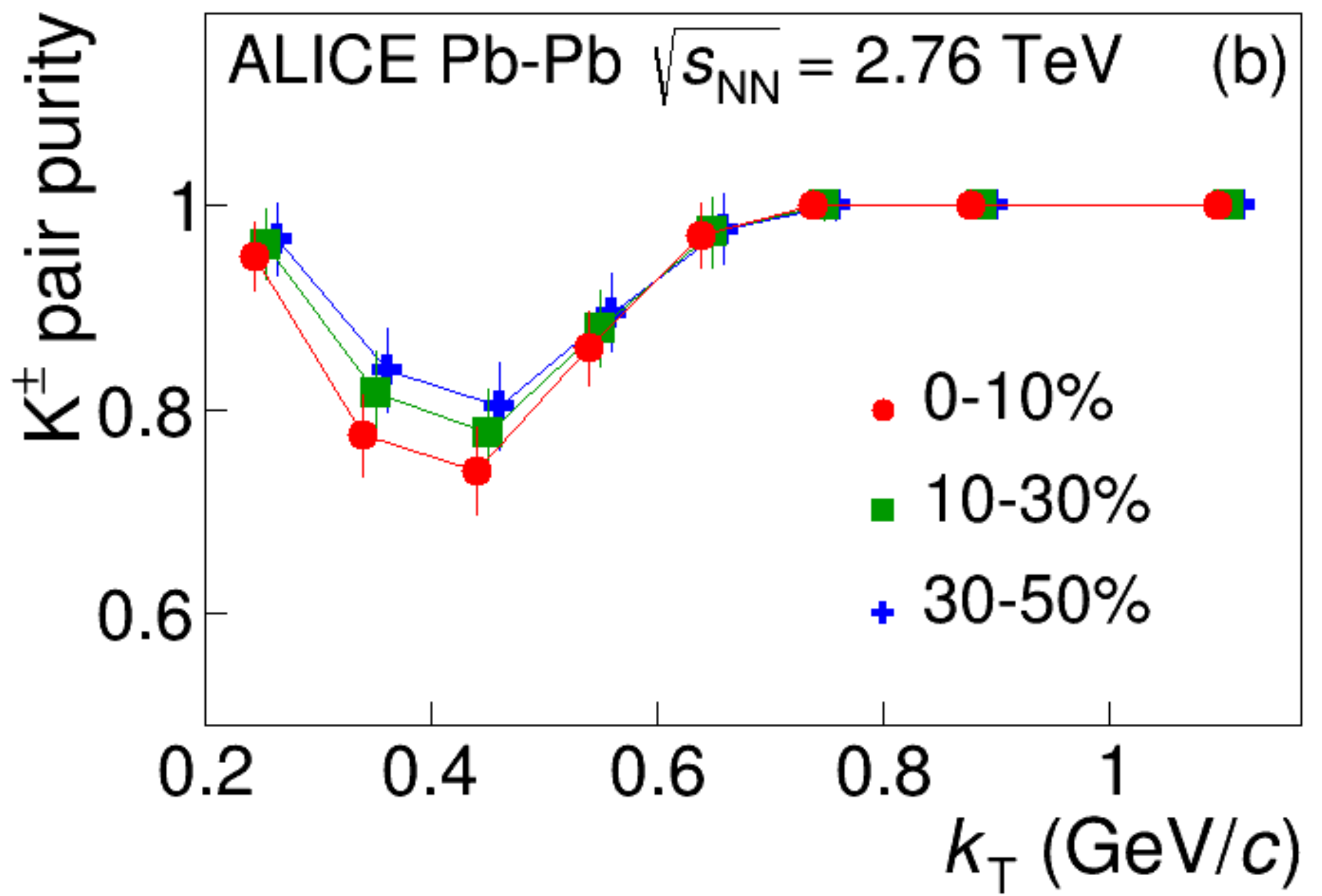}
    \caption{Single $\Kpm$ purity (a) and pair purity for small relative momenta (b) for different centralities. In (b) the $k_{\rm T}$ values for different centrality intervals are slightly offset for clarity.}
    \label{fig1}
  \end{center}
\end{figure}

$\Kpm$ identification was performed using the TPC (for all momenta) and the TOF detector (for $p > 0.5$ GeV/$c$).
The use of different values for $N_{\sigma,{\mathrm{TPC}}}$ and $N_{\sigma,{\mathrm{TOF}}}$ was
the result of studies to obtain the best kaon purity,
defined as the fraction of accepted kaon tracks that correspond to true kaon particles, while retaining a decent efficiency.
The estimation of purity for $p < 0.5$~GeV/c was performed by parametrizing the TPC d$E$/d$x$ distribution in momentum slices for the contributing species
\cite{Abelev:2014ffa}.
The dominant contamination for charged kaons comes from $\rm e^{\pm}$ in the
momentum range $0.4<p<0.5$~GeV/$c$.
The purity for $p>0.5$~GeV/$c$, where the TOF information was
employed, was studied with HIJING~\cite{Wang:1991hta} simulations using GEANT~\cite{Brun:1994aa} to model particle transport through the detector;
the charged kaon purity was estimated to be greater than 99\%.
The momentum dependence of the single kaon purity is shown
in Fig.~\ref{fig1}(a). 
The pair purity is calculated as the product of two single-particle purities,
where the momenta are taken from the experimentally determined distribution. The $\Kpm$ pair purity as a
function of $k_{\rm T}$ at three different centralities is shown in Fig.\ref{fig1}(b). Kaon pair transverse momentum is an averaged $p_{\rm T}$ of single kaons taken from the whole $p_{\rm T}$ range, which is the reason why the pair purities are larger than single particle ones. 

\FloatBarrier

Two kinds of two-track effects have been investigated: splitting, where a signal produced by one particle is incorrectly reconstructed as two tracks, and merging, where two particles are reconstructed as only one track.
These detector inefficiencies can be suppressed by employing specific pair selection criteria. We used the same procedure as in \cite{Adam:2015vna} which works here as well with slightly modified cut values.
Charged kaon pairs were required to have a separation of \mbox{$|\Delta \varphi^{*}|>0.04$} and \mbox{$|\Delta \eta|>0.02$}. Here, $\varphi^{*}$ is the azimuthal position of the track in the TPC at $R$ = 1.2 m, taking into account track curvature in the magnetic field, and $\eta$ is the pseudorapidity.
Also, all track pairs sharing more than $5 \%$ of TPC clusters were rejected.

\begin{table}
\centering
\begin{tabular}{l|l}
  \hline
  \multicolumn{2}{c}{Charged kaon selection} \\ \hline
    $p_{\rm T}$  & $0.15<p_{\rm T}<1.5$ GeV/$c$ \\ \hline
    $|\eta|$ & $< 0.8$ \\ \hline
    $\rm DCA_{\rm transverse}$ to primary vertex & $< 2.4$ cm \\ \hline
    $\rm DCA_{\rm longitudinal}$ to primary vertex & $< 3.0$ cm \\ \hline
    $N_{\sigma,\rm TPC}$ (for $p < 0.5$~GeV/$c$) & $< 2$ \\ \hline
    $N_{\sigma,\rm TPC}$ (for $p > 0.5$~GeV/$c$) & $< 3$ \\ \hline
    $N_{\sigma,\rm TOF}$ (for $0.5 < p < 0.8$ GeV/$c$) & $< 2$ \\ \hline
    $N_{\sigma,\rm TOF}$ (for $0.8 < p < 1.0$ GeV/$c$) & $< 1.5$ \\ \hline
    $N_{\sigma,\rm TOF}$ (for $1.0 < p < 1.5$ GeV/$c$) & $< 1.0$ \\ \hline
  \hline
  \multicolumn{2}{c}{Neutral kaon selection} \\ \hline
    $|\eta|$ & $< 0.8$ \\ \hline
    Daughter-daughter $\rm DCA_{3D}$ & $< 0.3$ cm \\ \hline
    $\rm DCA_{3D}$ to primary vertex & $< 0.3$ cm \\ \hline
    Invariant mass & $0.480 < m_{\pi^+ \pi^-} < 0.515$ GeV/${c}^2$ \\ \hline
    Daughter $p_{\rm T}$ & $> 0.15$ GeV/$c$ \\ \hline
    Daughter $|\eta|$ & $< 0.8$ \\ \hline
    Daughter $\rm DCA_{3D}$ to primary vertex & $> 0.4$ cm \\ \hline
    Daughter $N_{\sigma,\rm TPC}$ & $< 3$ \\ \hline
    Daughter $N_{\sigma,\rm TOF}$ (for $p > 0.8$ GeV/$c$) & $< 3$ \\ \hline
\end{tabular}
\caption{Single particle selection criteria.}
\label{tab:K0cuts}
\end{table}

\subsection{Neutral kaon selection}

The decay channel $\Kzs\rightarrow\pi^+\pi^-$ was used for the identification of neutral kaons. The secondary pion tracks were reconstructed using TPC and ITS information.
 The single-particle cuts for parents ($\Kzs$) and daughters ($\pi^{\pm}$) used in the decay-vertex reconstruction are shown in Table \ref{tab:K0cuts}.
The daughter-daughter DCA, that is the distance of closest approach of the two daughter pions from a candidate $\Kzs$ decay, proved useful in rejecting background topologies.
PID for the pion daughters was performed using both TPC (for all momenta) and TOF (for $p>0.8$~GeV/$c$). 
The very good detector performance is reflected in the FWHM of the $\Kzs$ peak of only 8~MeV/$c^2$.
The selection criteria used in this analysis were chosen as a compromise to maximize statistics while keeping a high signal purity.
 The neutral kaon purity (defined as Sig./[Sig.+Bkg.]\ for $0.480 < m_{\pi^+ \pi^-} < 0.515$~GeV/$c^2$) was larger than 0.95.

Two main two-particle cuts were used in the neutral kaon analysis. To resolve two-track inefficiencies associated with the daughter tracks, such as the splitting or merging of tracks discussed above,
a separation cut was employed in the following way. For each kaon pair, the spatial separation between the same-sign pion daughters
was calculated at several points throughout the TPC (every 20 cm radially from 85 cm to 245 cm) and averaged.
If the average separation of either pair of tracks was below 5 cm, the kaon pair was not used. Another cut was used to prevent two reconstructed kaons from using the same daughter track. If two kaons shared a daughter track,
one of them was excluded using a procedure which compared the two $\Kzs$ candidates and kept the candidate
whose reconstructed parameters best matched those expected for a true $\Kzs$ particle in two of three categories
(smaller $\Kzs$ DCA to primary vertex, smaller daughter-daughter DCA, and $\Kzs$ mass closer to the PDG value~\cite{Agashe:2014kda}).
This procedure was shown, using HIJING+GEANT simulations, to have a success rate of about 95\% in selecting a true $\Kzs$ particle over a fake one.
More details about the $\Kzs\Kzs$ analysis can be found in Refs.~\cite{Matt_thesis,Abelev:2012ms}. $\Kzs$ candidate selection criteria developed in other works \cite{Adam:2015vja} were used here as well; they are included in Table \ref{tab:K0cuts}.

\section{\label{subsec11}Correlation functions}

The femtoscopic correlation function $C$ is constructed experimentally as
the ratio $C({\bf q}) = A({\bf q})/B({\bf q})$, where $A({\bf q})$ is the measured distribution of the difference ${\bf q} = {\bf p}_{2}-{\bf p}_{1}$ between the three-momenta of the two particles  ${\bf p_1}$ and ${\bf p_2}$ taken from the same event, $B({\bf q})$ is a reference
distribution of pairs of particles taken from different events (mixed).
For a detailed description of the formalism, see e.g. \cite{Lednicky:2005af}.
The pairs in the denominator distribution $B(\textbf{q})$ are constructed by taking a particle from one event and pairing it with a particle from another event with a similar centrality and primary vertex position along the beam direction. Each event is mixed with five (ten) others for the $\Kzs$ ($\Kpm$) analysis.
The numerator and denominator are normalized in the full
$q = \sqrt{|{\bf q}|^{2} - q_{0}^{2}}$ range used
(0--0.3~GeV/$c$) such that $C(q) \rightarrow 1$ means no correlation.
Pair cuts have been applied in exactly the same way for the same-event (signal) and mixed-event (background) pairs.

 The momentum difference is calculated in the longitudinally co-moving system (LCMS), where the longitudinal pair momentum vanishes, and is decomposed into ($q_{\rm out}$, $q_{\rm side}$, $q_{\rm long}$), with the ``long'' axis going  along the beam, ``out'' along the pair transverse momentum, and ``side'' perpendicular to the latter in the transverse plane (Bertsch-Pratt convention).

The correlation functions have been corrected for momentum resolution effects, by using the HIJING event generator and assigning a quantum-statistical weight to each particle pair. Further, these modified events were propagated through the full simulation of the ALICE detectors \cite{Aamodt:2008zz}. The ratios  of the correlation functions obtained before and after this full event simulation have been taken as the correction factors.
The correlation function from the data has been divided by this $q$-dependent factor. The correction increases the obtained radii by 3--5\%.

\subsection{Charged kaon}
The three-dimensional correlation functions were fitted by the Bowler-Sinyukov formula ~\cite{Bowler:1991vx,Sinyukov:1998fc}:
\begin{equation}
C({\bf q}) = N \left(1 -\lambda \right) + N\lambda K(q)\left[ 1+
\exp \left(-R_{\rm out}^{2} q_{\rm out}^{2}-R_{\rm side}^{2} q_{\rm side}^{2}-R_{\rm long}^{2} q_{\rm long}^{2}\right)\right],
\label{eq:CF}
\end{equation}
where $R_{\rm out}$, $R_{\rm side}$, and $R_{\rm long}$ are the Gaussian femtoscopic radii in the LCMS frame, $N$ is the normalization factor, and $q$ is the momentum difference in the pair rest frame (PRF) \footnote{Average $q$ in PRF for the given ``out-side-long'' bin is determined during the $C({\bf q})$ construction and used as an argument of the $K$-function.}
. The $\lambda$  parameter, which characterizes the correlation strength, can be affected
 by long-lived resonances, coherent sources~\cite{Akkelin:2001nd,Abelev:2013pqa,Wiedemann:1996ig}, and non-Gaussian features of the particle-emission distribution.
We account for Coulomb effects through $K(q)$, calculated according to Ref.~\cite{Sinyukov:1998fc,Abelev:2013pqa} as
\begin{equation}
K(q) = C({\rm QS+ Coulomb}) / C(\rm QS). \label{eq:newCoulomb}
\end{equation}

\begin{figure}[!]
\begin{center}
\includegraphics[width=0.8\textwidth]{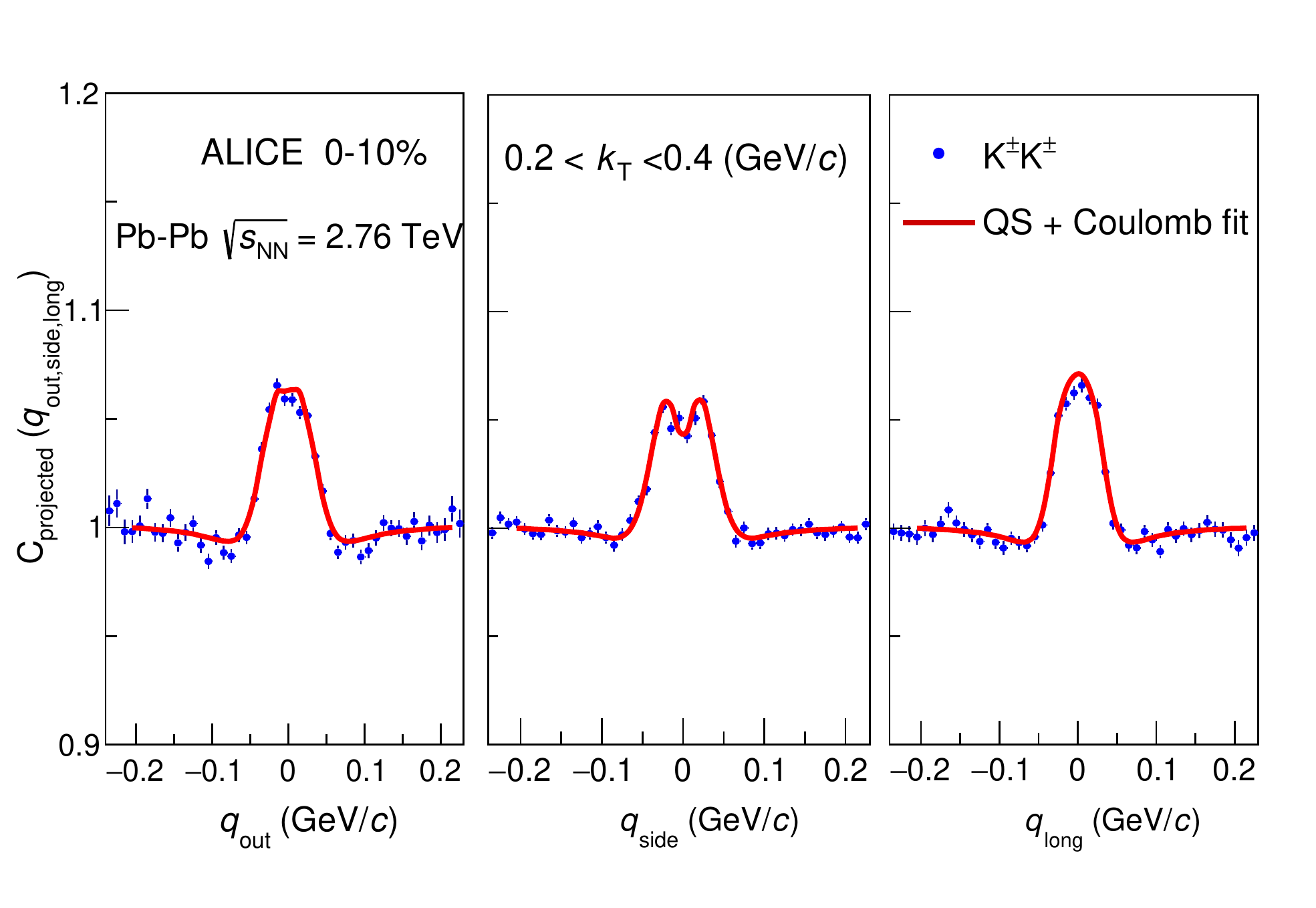}
\caption{A sample projected $\Kpm\Kpm$ correlation function with fit. The error bars are statistical only. Systematic uncertainties on the points are equal to or less than the statistical error bars shown.}
\label{figCFKK}
\end{center}
\end{figure}
\FloatBarrier

Here, the theoretical correlation function $C(\rm QS)$ takes into account quantum statistics only and $C(\rm QS+\rm Coulomb)$  considers quantum statistics and the Coulomb final-state interaction (FSI) contribution to the wave function \cite{Lednicky:2005af}.

The experimental correlation functions have been corrected for purity according to:
\begin{equation}
C_{\mathrm{corrected}} = (C_{\mathrm{raw}}-1+\zeta)/\zeta,
\label{eq:puritycorrection}
\end{equation}
where $\zeta$ is the pair purity taken from
Fig.~\ref{fig1}.

Figure~\ref{figCFKK} shows a sample projected $\Kpm\Kpm$ correlation function with a fit performed according to Eq.~(\ref{eq:CF}). When the 3D correlation function is projected on one axis, the momentum differences in the two other directions are required to be within (-0.04,0.04)~GeV/$c$.

\subsection{Neutral kaon}
$\Kzs\Kzs$ correlation functions were fitted using a parametrization which includes Bose-Einstein statistics as well as strong final-state interactions~\cite{Lednicky:1981su,Abelev:2006gu}.
Strong final-state interactions have an important effect on $\Kzs\Kzs$ correlations. Particularly, the $\kz\kzb$ channel is affected by the near-threshold resonances $\mathrm{f}_0(980)$ and $\mathrm{a}_0(980)$.  Using the equal emission time approximation in the pair rest frame (PRF)~\cite{Lednicky:1981su}, the elastic $\kz\kzb$ transition is written as a stationary solution $\Psi_{-\vkstar}(\vrstar)$ of the scattering problem in the PRF,
where $\vkstar$ and $\vrstar$ represent the momentum of a particle and the emission separation of the pair in the PRF (the $-\vkstar$ subscript refers to a reversal of time from the emission process), which at large distances has the asymptotic form of a superposition of a plane wave and an outgoing spherical wave,
\begin{equation}
\Psi_{-\vkstar}(\vrstar) = e^{-i\vkstar \cdot \vrstar} + g(\kstar) \dfrac{e^{i\kstar\rstar}}{\rstar} \;,
\label{eq:FSIwave}
\end{equation}
where $g(\kstar)$ is the s-wave scattering amplitude for a given system. For $\kz\kzb$, $g(\kstar)$ is dominated by the $\mathrm{f}_0$ and $\mathrm{a}_0$ resonances and written in terms of the resonance masses and decay couplings~\cite{Abelev:2006gu}:
\begin{align}
g(\kstar) &= \dfrac{1}{2} [ g_0(\kstar) + g_1(\kstar)] \;,\\
g_I(\kstar) &= \dfrac{\gamma_r}{m_r^2-s-i\gamma_r\kstar-i\gamma_r'k_r'} \; .
\label{eq:ScatteringAmps}
\end{align}
Here, $s = 4(m^2_K+k^{*2})$; $\gamma_r (\gamma_r')$ refers to the couplings of the resonances to the $\mathrm{f}_0 \rightarrow \kz\kzb (\mathrm{f}_0 \rightarrow \pi\pi)$ and $\mathrm{a}_0 \rightarrow \kz\kzb (\mathrm{a}_0 \rightarrow \pi\eta)$ channels; $m_r$ is the resonance mass; and $k_r'$ refers to the momentum in the PRF of the second decay channel ($\mathrm{f}_0 \rightarrow \pi\pi$ or $\mathrm{a}_0 \rightarrow \pi\eta$) with the corresponding partial width $\Gamma_r' = \gamma_r' k_r'/m_r$\,. The amplitudes $g_I$ of isospin $I=0$ and $I=1$ refer to the $\mathrm{f}_0$ and $\mathrm{a}_0$, respectively. The parameters associated with the resonances and their decays are taken from several experiments \cite{Antonelli:2002ip,Achasov:2001cj,Achasov:2002ir,Martin:1976vx}, and the values are listed in Table~\ref{tab:f0a0}.

\begin{table}
 \centering
 \begin{tabular}{| l | l | l | l | l | l | l |}
  \hline
  Ref & $m_{\mathrm{f}_0}$ & $\gamma_{\mathrm{f}_0K\bar{K}}$ & $\gamma_{\mathrm{f}_0\pi\pi}$ & $m_{\mathrm{a}_0}$ & $\gamma_{\mathrm{a}_0K\bar{K}}$ & $\gamma_{\mathrm{a}_0\pi\eta}$ \\ \hline
  \cite{Antonelli:2002ip} & 0.973 & 2.763 & 0.5283 & 0.985 & 0.4038 & 0.3711 \\ \hline
  \cite{Achasov:2001cj} & 0.996 & 1.305 & 0.2684 & 0.992 & 0.5555 & 0.4401 \\ \hline
  \cite{Achasov:2002ir} & 0.996 & 1.305 & 0.2684 & 1.003 & 0.8365 & 0.4580 \\ \hline
  \cite{Martin:1976vx} & 0.978 & 0.792 & 0.1990 & 0.974 & 0.3330 & 0.2220 \\
  \hline
  \end{tabular}
  \caption[$\mathrm{f}_0$ and $\mathrm{a}_0$ masses and coupling parameters]{The $\mathrm{f}_0$ and $\mathrm{a}_0$ masses and coupling parameters, all in GeV.}
  \label{tab:f0a0}
\end{table}

The correlation function is then calculated by integrating $\Psi_{-\vkstar}(\vrstar)$ in the Koonin-Pratt equation~\cite{Koonin:1977fh,Pratt:1990zq}
\begin{equation}
C(\vkstar,\vec{K}) = \int d^3 \, \vrstar \, S_{\vec{K}}(\vrstar) \lvert \Psi^S_{-\vkstar}(\vrstar) \rvert ^2 \, ,
\label{eq:KooninPratt}
\end{equation}
where $S_{\vec{K}}(\vrstar)$ is the Gaussian source distribution in terms of $R_{\rm out}$,
$R_{\rm side}$, and $R_{\rm long}$, $\vec{K}$ is the average pair momentum, and
$\Psi^S_{-\vkstar}(\vrstar)$ is the symmetrized version of $\Psi_{-\vkstar}(\vrstar)$ for bosons.
Although Eq.~(\ref{eq:KooninPratt}) can be integrated analytically for $\kzs\kzs$ correlations
with FSI for the one-dimensional case \cite{Abelev:2006gu}, for the three-dimensional case
this integration cannot be performed analytically. In order to form the 3D correlation function, we combine a Monte Carlo emission simulation with a calculation of the two-particle wavefunction, thus performing a numerical integration of Eq.~(\ref{eq:KooninPratt}). The Monte Carlo emission simulation consists of generating the pair positions sampled from a three-dimensional Gaussian in the PRF, with three input radii as the width parameters, and generating the particle momenta sampled from a distribution taken from data. Using the MC-sampled positions and momenta, we calculate $\Psi^S_{-\vkstar}(\vrstar)$.
We then build a correlation function using the wavefunction weights to form the signal distribution, and an unweighted distribution acts as a background.
This theoretical correlation function is then used to fit the data. Finally, we make a Lorentz boost,
$\gamma$, of $R_{\rm out}$ from the PRF to the LCMS frame ($R_{\rm side}$ and $R_{\rm long}$
are not affected by the boost).
More details on the 3D fitting procedure can be found in Ref.~\cite{Matt_thesis}.

\begin{figure}[t!]
\begin{center}
\includegraphics[width=1.0\textwidth]{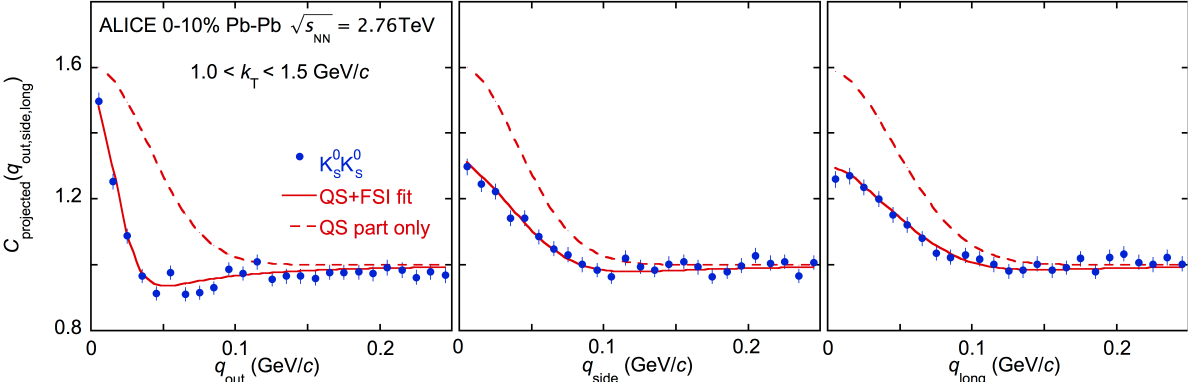}
\caption{A sample projected $\Kzs\Kzs$ correlation function with fit. Also shown is the contribution to the fit from the quantum statistics part only. The error bars are statistical only. Systematic uncertainties on the points are equal to or less than the statistical error bars shown.}
\label{fig2}
\end{center}
\end{figure}

Figure~\ref{fig2} shows a sample projected $\Kzs\Kzs$ correlation function with fit. Also shown is the contribution to the fit from the quantum statistics part only. As seen, the FSI part produces a significant depletion of the correlation function in the $q$ range 0--0.1~GeV/$c$ in each case.
\FloatBarrier

\subsection{Systematic uncertainties}

The effects of various sources of systematic uncertainty on the extracted fit parameters were studied as functions of centrality and $k_{\rm T}$.  For each source, we take the maximal deviation and apply it symmetrically as the uncertainty.
Table~\ref{tab:systerrKch} shows minimum and maximum uncertainty values for various sources of systematic uncertainty for charged and neutral kaons. The systematic errors are summed up quadratically.
The values of the total uncertainty are not necessarily equal to the sum of the  individual uncertainties, as the latter can come from different centrality or $k_{\rm T}$ bins.
Both analyses studied the effects of changing the selection criteria used for the events, particles, and pairs (variation of cut values up to $\pm$25$\%$) and varying the range of $q$ values over which the fit is performed (variation of $q$ limits up to $\pm$25$\%$). Uncertainties associated with momentum resolution corrections are included into the $\Kpm$ analysis; for the $\Kzs$ analysis, these uncertainties are found to be small compared to other contributions.
Both analyses were performed separately for the two different polarities of the ALICE solenoid magnetic field, the difference was found to be negligible.

For the $\Kzs$ fitting procedure, the mean $\gamma$ value
is calculated for each centrality and $k_{\rm T}$ selection and used to
scale $R_{\rm out}$. However, each bin has a spread of $\gamma$ values associated with it.
The standard deviation of the mean $\gamma$ value for each $k_{\rm T}$  bin was used as an additional source of systematic error for $R_{\rm out}$.
 For $\Kzs$, an uncertainty on the strong FSI comes from the fact that several sets of $\rm \mathrm{f}_0(980)$ and $\rm \mathrm{a}_0(980)$ parameters are available~\cite{Martin:1976vx,Antonelli:2002ip,Achasov:2001cj,Achasov:2002ir}; each set is used to fit the data, the results are averaged, and the maximal difference was taken as a systematic error.

 The $\Kpm$ analysis has uncertainties associated with the
choice of the radius for the Coulomb function.
For each correlation function it is set to the value from the one-dimensional analysis \cite{Adam:2015vja}.
Its variation by $\pm 1$~fm is a source of systematic uncertainty.
Another source of systematic uncertainty is misidentification of particles and the associated purity correction. A 10\% variation of the parameters in the purity correction was performed.
 We also incorporated sets with a reduced electron contamination by
 I) tightening the PID criteria, in particular extending the momentum range where the TOF signal was used and requiring the energy-loss measurement to be consistent with the kaon hypothesis within one sigma, and II) completely excluding the momentum range 0.4--0.5~GeV/$c$.

\begin{table}
 \centering
 \begin{tabular}{cc|c|c|c|c}
   \multicolumn{1}{c}{} & $R_{\rm out}$ [\%] & $R_{\rm side}$ [\%] & $R_{\rm long}$  [\%] & $\lambda$ [\%] \\ \hline
\hline
  \multicolumn{2}{c}{Charged kaon} \\ \hline
   \multicolumn{1}{l}{Single particle selection} & 0--2 & 0--2 &  0--2 & 0--2  \\ \hline
   \multicolumn{1}{l}{PID and purity} & $<$0.1 &  $<$0.1 & $<$0.1 & 1--10   \\ \hline
   \multicolumn{1}{l}{Pair selection} & 2--8 & 1--6 & 2--10 & 6--15 \\ \hline
   \multicolumn{1}{l}{Fit range} & 1--3 & 1--4 & 1--7 & 1--7 \\ \hline
   \multicolumn{1}{l}{Coulomb function} & 3--5 &  1--2 & 2--3 & 8--10 \\ \hline
   \multicolumn{1}{l}{Momentum resolution} & 1--2 & 1--2 & 1--3 & 2--6 \\ \hline
   \multicolumn{1}{l}{\bf Total (quad. sum)} & 7--11 & 7--9 & 7--12 & 10--17 \\ \hline
  \hline
  \multicolumn{2}{c}{Neutral kaon} \\ \hline
  \multicolumn{1}{l}{Single particle and pair selection} & 0--1 & 1--5 &  1--4 & 6--14  \\ \hline
   \multicolumn{1}{l}{Pair selection} & 2--8 & 1--6 & 2--10 & 6--15 \\ \hline
   \multicolumn{1}{l}{FSI Model} & 1--6 & 1--6 & 1--15 & 3--9   \\ \hline
   \multicolumn{1}{l}{$\gamma$} & 5--10 & $<$0.1 & $<$0.1 & $<$0.1 \\ \hline
   \multicolumn{1}{l}{Fit range} & 0--6 & 0--6 & 0--10 & 0--6 \\ \hline
   \multicolumn{1}{l}{Momentum resolution} & $<$0.1 & 0--3 & 0--6 & 2--3 \\ \hline
   \multicolumn{1}{l}{\bf Total (quad. sum)} &6--11 & 3--7 & 2--15 & 7--16 \\ \hline
 \end{tabular}
 \caption{Minimum and maximum uncertainty values for various sources of systematic uncertainty for charged and neutral kaons (in percent).
 Note that each value is the maximum uncertainty from a specific source,
  but can pertain to a different centrality or $k_{\rm T}$ bin.
 Thus, the maximum total uncertainties are smaller than (or equal to) the quadratic sum of the maximum individual uncertainties.}
 \label{tab:systerrKch}
\end{table}

\section{Results and discussion}
\label{res}

\begin{figure}[t!]
\begin{center}
\includegraphics[width=0.85\textwidth]{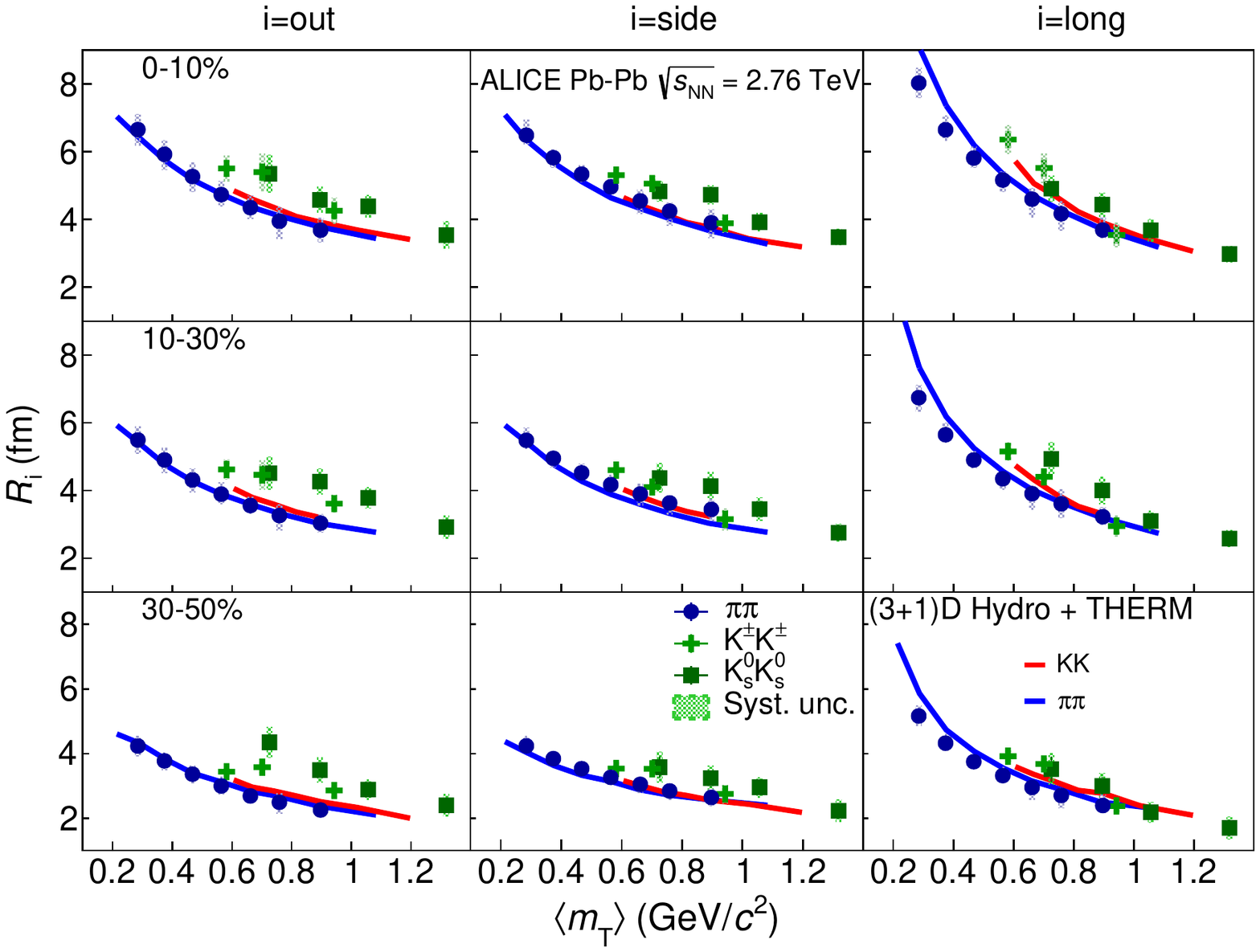}
\caption{ The 3D LCMS radii vs. $m_{\mathrm{T}}$  for charged (light green crosses) and neutral (dark green squares) kaons and pions \cite{Adam:2015vna} (blue circles) in comparison with
the theoretical predictions of the (3+1)D Hydro + THERMINATOR-2 model \cite{Kisiel:2014upa} for pions (blue solid lines) and kaons (red solid lines). }
 \label{fig:radii_THERM}
\end{center}
\end{figure}

Figure~\ref{fig:radii_THERM} shows the $m_{\mathrm{T}}$-dependence of the extracted femtoscopic radii $R_{\rm out}$, $R_{\rm side}$, and $R_{\rm long}$ in three centrality selections for pions \cite{Adam:2015vna} and charged and neutral kaons.
The obtained radii are smaller for more peripheral collisions than for central ones. The radii decrease with increasing $m_{\mathrm{T}}$ and each particle species roughly follows an $m_{\mathrm{T}}^{-1/2}$ dependence. The radii in ``out'' and ``long'' directions exhibit larger values for kaons than for pions at the same transverse mass demonstrating that the $m_{\mathrm{T}}$-scaling is broken.
This difference increases with centrality and is maximal for the most central collisions. 
Also presented in Fig.~\ref{fig:radii_THERM} are the predictions of the (3+1)D hydrodynamical model coupled with the statistical hadronization code THERMINATOR-2 \cite{Kisiel:2014upa}. The model describes well the $m_{\mathrm{T}}$-dependence of pion radii, but underestimates kaon radii. Consistent with the data, the (3+1)D Hydro+THERMINATOR-2 model shows mild breaking in the ``long'' direction for central collisions, but it underestimates the breaking in the ``out'' direction.
The  significance of this breaking of the scaling is discussed further in this section.

\begin{figure}[t!]
\begin{center}
\includegraphics[width=0.8\textwidth]{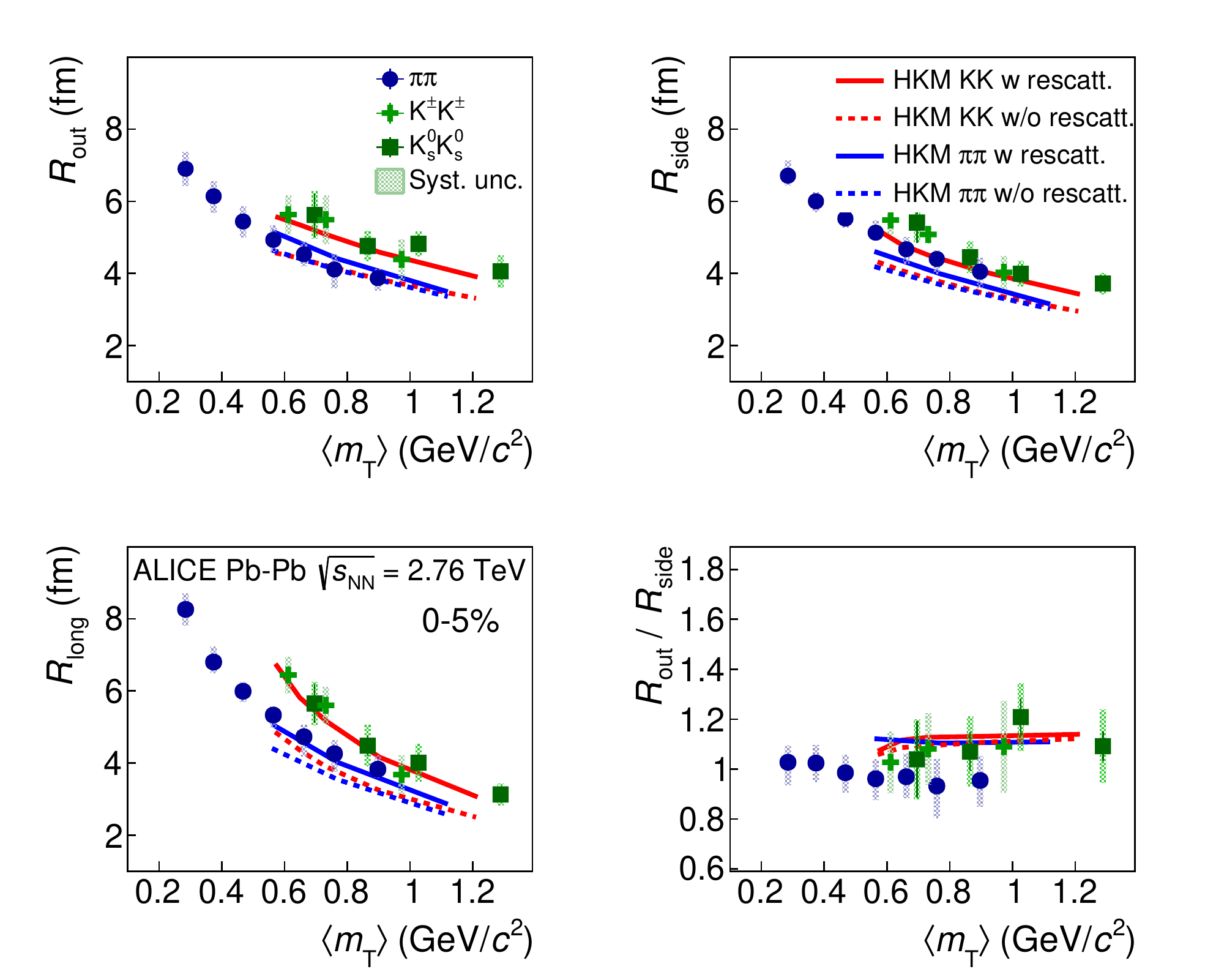}
\caption{ The 3D LCMS radii vs. $m_{\mathrm{T}}$
for 0--5\% most central collisions
in comparison with the theoretical predictions of HKM \cite{Shapoval:2014wya} for pions (blue lines) and kaons (red lines).
}
 \label{fig:radii_HKM}
\end{center}
\end{figure}

In addition to the aforementioned three-dimensional radii, here for
the 0--5\% most central events, Fig.~\ref {fig:radii_HKM} also shows
the $m_{\mathrm{T}}$ dependence of the ratio $R_{\rm out}/R_{\rm side}$ for charged and neutral kaons in comparison with HKM predictions \cite{Shapoval:2014wya} with and without the hadronic rescattering phase. The HKM calculations without rescattering exhibit an approximate $m_{\mathrm{T}}$-scaling but do not describe the data, while the data are well reproduced by the full hydro-kinetic model calculations thereby showing the importance of the rescattering phase at LHC energies.
The $R_{\rm out}$ and $R_{\rm side}$ radii are both influenced  by flow and rescatterings, so their ratio is rather robust against these effects. The fact that
$R_{\rm out}/R_{\rm side}$ ratio of pions and kaons coincide in the
HKM simulations (Fig.~\ref {fig:radii_HKM}) is related to some
underestimation of $R_{\rm side}$ radii for pions while pion $R_{\rm out}$ radii are
slightly overestimated in the model.

\FloatBarrier

\FloatBarrier
\begin{figure}[t!]
\begin{center}
\includegraphics[width=0.85\textwidth]{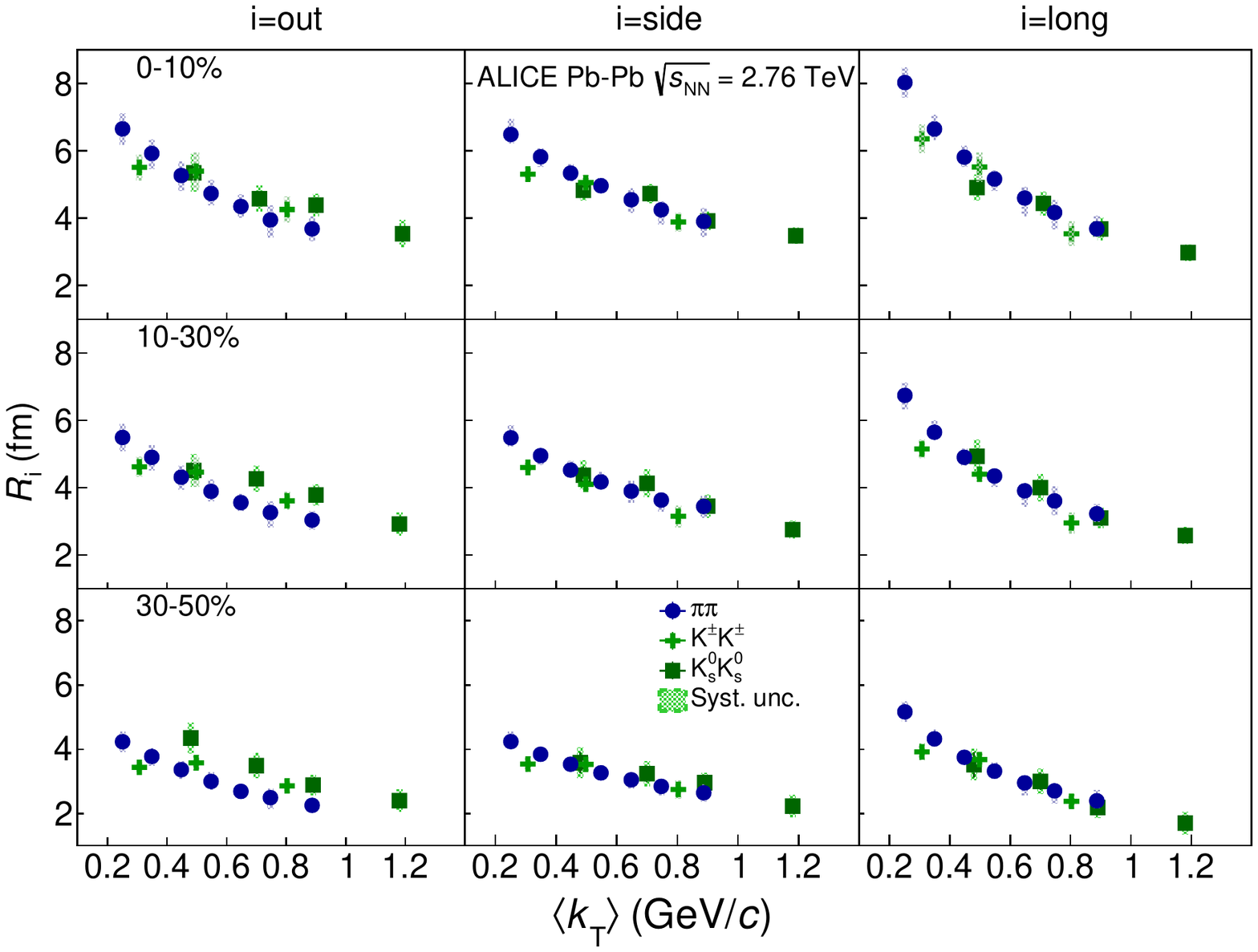}
\caption{ The 3D LCMS radii vs. $k_{\mathrm{T}}$  for charged (light green crosses) and neutral (dark green squares) kaons and pions \cite{Adam:2015vna} (blue circles).
}
 \label{fig:radii_kT}
\end{center}
\end{figure}
\FloatBarrier

 It was predicted in \cite{Shapoval:2014wya} that the radii scale better with $k_{\mathrm{T}}$ at LHC energies as a result of the interplay of different factors in the model, including the particular initial conditions.
Figure~\ref{fig:radii_kT} illustrates the $k_{\mathrm{T}}$-dependence of the femtoscopic radii $R_{\rm out}$, $R_{\rm side}$, and $R_{\rm long}$. Unlike the $m_{\mathrm{T}}$-dependence, the radii seem to scale better with $k_{\mathrm{T}}$
in accordance with this prediction.

\begin{figure}[t!]
\begin{center}
\includegraphics[width=0.9\textwidth]{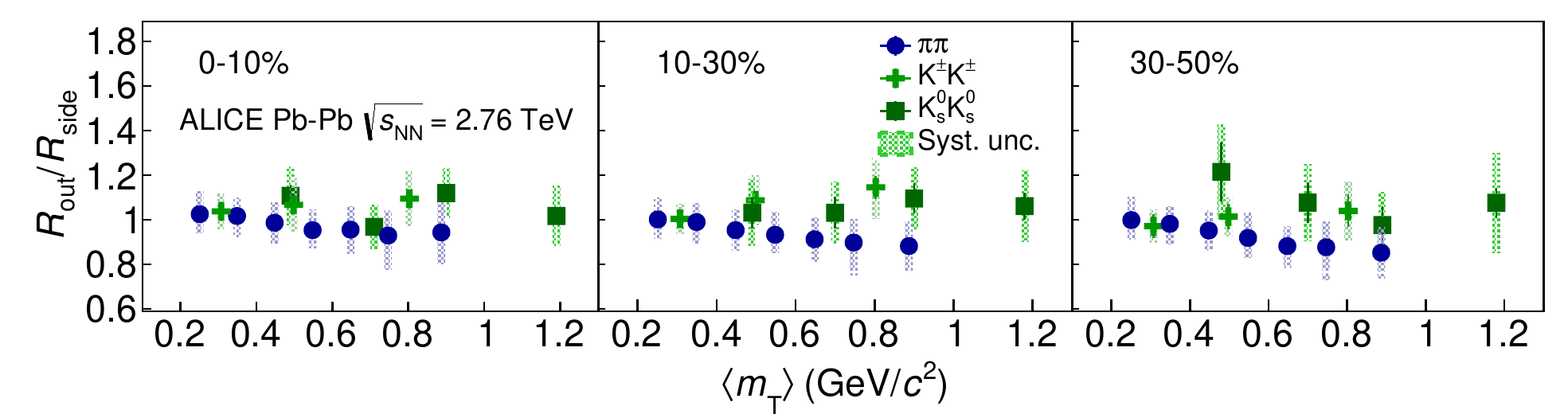}
\caption{ $R_{\rm out}/R_{\rm side}$  vs. $m_{\mathrm{T}}$ for pions \cite{Adam:2015vna} and kaons
for different centrality intervals.}
 \label{fig:Ros}
\end{center}
\end{figure}

 The ratio $R_{\rm out}/R_{\rm side}$ appears to be sensitive to the space-time correlations present at the freeze-out hypersurface \cite{Kisiel:2008ws,Karpenko:2009wf,Adam:2015vna}.
As it was observed in \cite{Adam:2015vna}, the ratio for pions is consistent with unity, slowly decreasing for more peripheral collisions and higher $k_{\mathrm{T}}$.
In Fig.~\ref{fig:Ros}, the ratio $R_{\rm out}/R_{\rm side}$ is shown for
pions and kaons at different centralities.
The systematic uncertainties partially cancel in the ratio. Systematic uncertainties are correlated in $m_{\rm T}$ for each type of particle pair; no correlation between the systematic uncertainties of the charged and neutral species exists.
The measured $R_{\rm out}/R_{\rm side}$ ratios are slightly larger for kaons than for pions. This is an indication of different space-time correlations for pions and kaons, and a more prolonged emission duration for kaons.

\begin{figure}[t!]
\begin{center}
\includegraphics[width=0.65\textwidth]{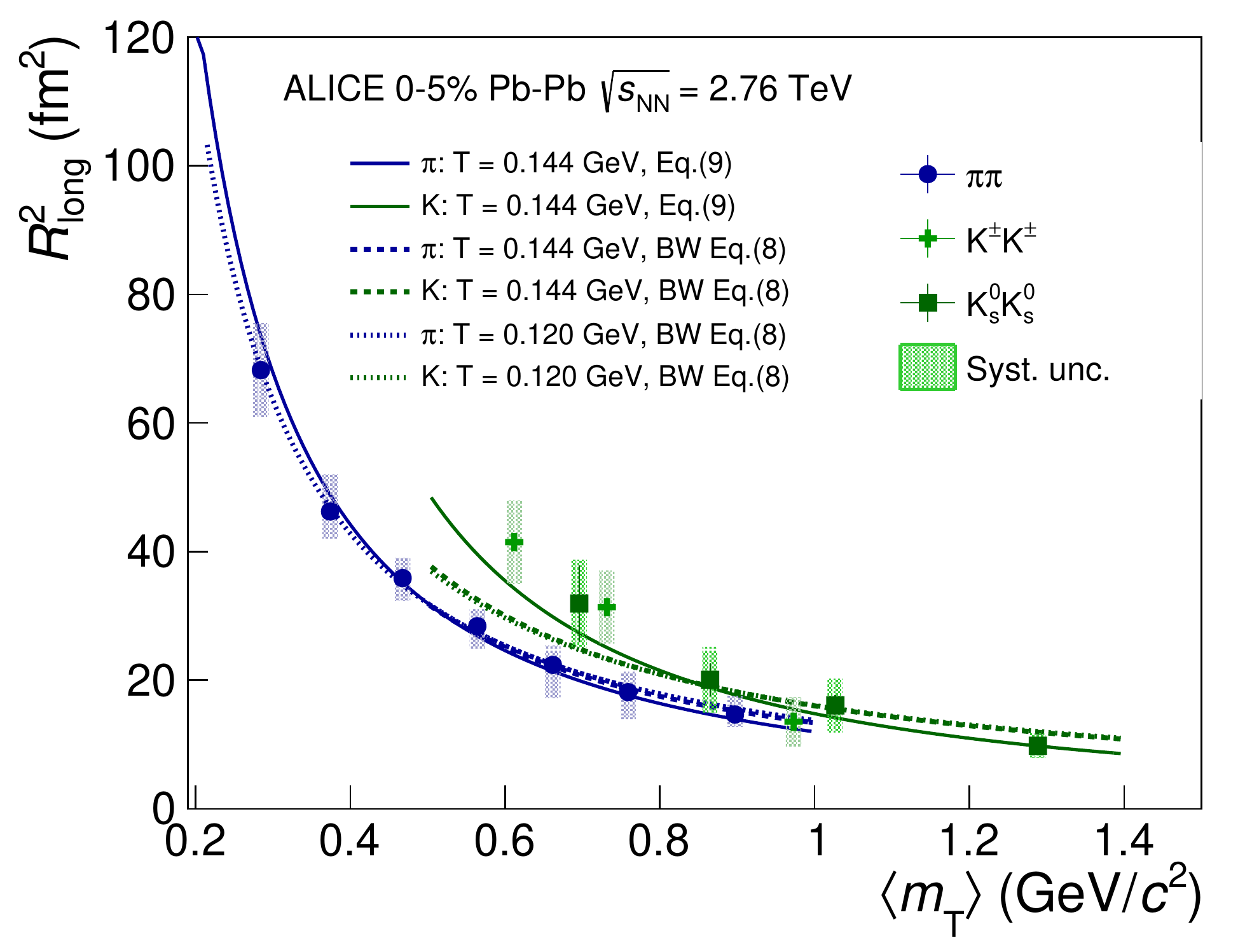}
\caption{  $R^{2}_{\rm long}$ vs. $m_{\mathrm{T}}$ for kaons and pions.
The solid lines show the fit using Eq.~(\ref{eq:tau_YS})
for pions and kaons to extract the emission times ($\tau$); the dashed and dotted lines
show the fit using Eq.~(\ref{eq:tau_stand}) with $T_{\rm kin}$ = 0.144~GeV and 
$T_{\rm kin}$ = 0.120~GeV, respectively. For pions at small $m_{\mathrm{T}}$, the dashed and dotted line coincide.
}
 \label{fig:tau}
\end{center}
\end{figure}

In our previous pion femtoscopy analysis \cite{Aamodt:2011mr} the information about the emission time (decoupling time) at kinetic freeze-out $\tau \sim 10$~fm/$c$
was extracted  by fitting the  $m_{\mathrm{T}}$-dependence of $R_{\rm long}^{2}$ using the blast-wave expression \cite{Herrmann:1994rr}:
\begin{equation}
R_{\rm long}^{2} = \tau^{2} \frac{T_{\rm kin}}{m_{\mathrm{T}}}\frac{K_{2}(m_{\mathrm{T}})}{K_{1}(m_{\mathrm{T}})},
\label{eq:tau_stand}
\end{equation}
where $T_{\rm kin}$ is the temperature at kinetic freeze-out, and $K_n$ are the integer-order modified Bessel functions.
We tried to use Eq.~(\ref{eq:tau_stand}) to fit the $R_{\rm long}^{2}$  $m_{\mathrm{T}}$-dependence (Fig.~\ref {fig:tau}) for pions and kaons
taking the thermal freeze-out temperature $T_{\rm kin}$ = 0.120~GeV as in \cite{Aamodt:2011mr}
(dotted lines) and $T_{\rm kin}$ = 0.144~GeV (dashed lines). The emission times extracted from the fit are presented in Table~\ref{tab:tau}.
However, though this formula works well for pions, it fails to describe kaon longitudinal radii. Large transverse flow may be partially responsible for this failure \cite{Sinyukov:2015kga}.
The following analytical formula for the time of maximal emission, $\tau_{max}$, is proposed in \cite{Sinyukov:2015kga}:
\begin{equation}
R_{\rm long}^{2} = \tau_{\rm max}^{2} \frac{T_{\rm max}}{m_{\mathrm{T}} \cosh{y_{\mathrm{T}}}}(1+\frac {3 T_{\rm max}} {2 m_{\mathrm{T}} \cosh{y_{\mathrm{T}}}} ),
\label{eq:tau_YS}
\end{equation}
where $\cosh {y_{\mathrm{T}}} = (1-v_{\mathrm{T}}^2)^{-1/2}$, $v_{\mathrm{T}} = \frac{\beta p_{\rm T}}{\beta m_{\mathrm{T}}+\alpha}$, $T_{\rm max}$ is the temperature at 
the hypersurface of maximal emission, $\beta = 1/T_{\rm max}$, and
$\alpha$ is a free parameter determining the intensity of flow \footnote{
The authors of \cite{Sinyukov:2015kga} use full evolutionary model (HKM) that has no sharp/sudden kinetic freeze-out. For such type of models a continuous  hadron emission takes place  instead. Then for each particle species, considered within certain transverse momentum bin, there is a 4D layer, adjacent to the space-like hypersurface of maximal emission, where most of selected particles are emitted from. This non-enclosed hypersurface is characterized by the (average) proper time $\tau_{\rm max}$ – time of maximal emission, and the effective temperature $T_{\rm max}$. The proposed  phenomenological expression  for $R_{\rm long}$ is associated just with this hypersurface and is based on the model that is different from the blast-wave parameterization for sudden freeze-out. So the blast-wave temperature  $T_{\rm kin}$ can differ from the temperature parameter $T_{\rm max}$.
}. 
The advantage of the formula given in Eq.~(\ref{eq:tau_YS}) is that it is derived for a scenario with transverse flow of any intensity, which is especially important for LHC energies.

The analytical formula Eq.~(\ref{eq:tau_YS}) was used to fit the $m_{\mathrm{T}}$-dependence of $R_{\rm long}^{2}$ (Fig.~\ref {fig:tau}).
The fit was performed using the following parameters determined in \cite{Sinyukov:2015kga} by fitting light flavor particle spectra \cite{Abelev:2013vea}: $T_{\rm max}$ = 0.144~GeV, and $\alpha_{\pi}$ = 5.0 and $ \alpha_{K}$ = 2.2.

The extracted times of maximal emission are presented in Table~\ref{tab:tau}.

\begin{table}
 \centering
 \begin{tabular}{| l | l | l | l | l | l |}
  \hline
  method               & $T$ (GeV) & $\alpha_{\pi}$ & $\alpha_K$ & $\tau_{\pi}$ (fm/$c$) & $\tau_{K}$ (fm/$c$) \\ \hline
  fit with BW Eq.~(\ref{eq:tau_stand}) & 0.120 & - & - &  9.6 $\pm$ 0.2 & 10.6 $\pm$ 0.1 \\ \hline
  fit with BW Eq.~(\ref{eq:tau_stand}) & 0.144 & - & - &  8.8 $\pm$ 0.2 & 9.5 $\pm$ 0.1  \\ \hline
 \hline
  fit with Eq.~(\ref{eq:tau_YS})    & 0.144 & 5.0   & 2.2   &  9.3 $\pm$ 0.2 & 11.0 $\pm$ 0.1  \\
  \hline
  fit with Eq.~(\ref{eq:tau_YS})    & 0.144 & 4.3 $\pm$ 2.3   & 1.6 $\pm$ 0.7   &  9.5 $\pm$ 0.2 & 11.6 $\pm$ 0.1 \\ \hline
  \end{tabular}
  \caption[emission times for pions and kaons]{Emission times for pions and kaons extracted using the Blast-wave formula
 Eq.~(\ref{eq:tau_stand}) and the analytical formula Eq.~(\ref{eq:tau_YS}).}
  \label{tab:tau}
\end{table}


 In order to estimate the systematic errors of the extracted times of maximal emission we also have performed fitting with $T_{\rm max}$, $\alpha_{\pi}$ and $\alpha_{K}$ varied within the range of their uncertainty \cite{Sinyukov:2015kga}: $\pm$0.03~GeV, $\pm$3.5 and $\pm$0.7, respectively. The maximum deviations from the central values appeared to be (+1.8, -0.5)~fm/$c$ for pions and  (+0.5, -0.1)~fm/$c$ for kaons.
These systematic errors are fully correlated. Regardless of the specific parameter choice, we consistently observe the time of maximal emission for kaons to be larger than the one for pions.
The extracted times of maximal emission are rather close to those obtained within the HKM model \cite{Sinyukov:2015kga}: $\tau_{\pi} = 9.44 \pm 0.02$~fm/$c$, $\tau_{K} = 12.40 \pm 0.04$~fm/$c$ \footnote{These results were obtained in \cite{Sinyukov:2015kga} using the small interval $q$=0-0.04~GeV/$c$ in order to minimize influence of the non-Gaussian tails. It is found in \cite{Sinyukov:2015kga} that if even strong non-Gaussian behavior is observed for the kaon correlation function in wide $q$-interval, one can nevertheless utilize the same formula Eq.~(\ref{eq:tau_YS}), but making free the parameter $\alpha$ for kaons. Then one gets practically the same effective time for kaon emission, as it is obtained from the fit of the correlation function in the small interval $q$=0-0.04~GeV/$c$; for pions there is no such problem.}.
There is evidence that the time of maximal emission for pions is smaller than the one for kaons. This observation can explain the observed breaking of $m_{\mathrm{T}}$-scaling between pions and kaons.
It is interesting to note that in \cite{Sinyukov:2015kga} this difference in the emission times is explained by the different influence of resonances on pions and kaons during the rescattering phase due to kaon rescattering through the $\rm {K}^{*}$(892) resonance (with lifetime of 4--5~fm/$c$).
 It was shown in \cite{Sinyukov:2015kga} that a significant regeneration of the $\rm{K}^{*}(892)$ takes place in full HKM simulations with rescatterings (UrQMD cascade), whereas this process is not present in a scenario where only resonance decays are taken into account.

Similar findings were reported in \cite{Knospe:2015nva}, where the production yield of $\rm{K}^*$(892) in heavy-ion collisions at the LHC was studied.
Also there, the inclusion of a hadronic phase in the theoretical modeling of the production process proved to be essential in order to reproduce the experimentally found suppression pattern of $\rm{K}^{*}(892)$ production when compared to pp
collisions \cite{Aamodt:2010cz}.

\section{Summary}
\label{sum}
We presented the first results of three-dimensional femtoscopic analyses
for charged and neutral kaons in Pb--Pb collisions at $\sqrt{s_{\rm{NN}}}$ = 2.76 TeV.

 A decrease of source radii with increasing transverse mass and decreasing event multiplicity was observed.
The $m_{\mathrm{T}}$ scaling expected by pure hydro-dynamical models appears to be broken in our data. A scaling of pion and kaon radii with $k_{\rm T}$ was observed instead. The measured ratio of transverse radii $R_{\rm out}/R_{\rm side}$ is larger for kaons than for pions, indicating different space-time correlations. A new approach \cite{Sinyukov:2015kga} for extracting the emission times for pions and especially for kaons was applied. It was shown that the measured time of maximal emission for kaons is larger than that of pions.

The comparison of measured three-dimensional radii with a model,
wherein the hydrodynamic phase is followed by the hadronic rescattering phase \cite{Shapoval:2014wya}, and pure hydrodynamical calculations \cite{Kisiel:2014upa,Shapoval:2014wya} has shown that pion femtoscopic radii are well reproduced by both approaches while the behavior of the three-dimensional kaon radii can be described only if the hadronic rescattering phase
is present in the model.

\newenvironment{acknowledgement}{\relax}{\relax}
\begin{acknowledgement}
\section*{Acknowledgements}
\input{fa_2017-07-19.tex}    
\end{acknowledgement}
\addcontentsline{toc}{section}{\bibname}
\bibliographystyle{utphys}
\bibliography{biblio}

\newpage
\appendix
\section{The ALICE Collaboration}
\label{app:collab}
\input{Alice_Authorlist_2017-Jul-19.tex}  
\end{document}

%% file: fa_2017-07-19.tex

The ALICE Collaboration would like to thank all its engineers and technicians for their invaluable contributions to the construction of the experiment and the CERN accelerator teams for the outstanding performance of the LHC complex.
The ALICE Collaboration gratefully acknowledges the resources and support provided by all Grid centres and the Worldwide LHC Computing Grid (WLCG) collaboration.
The ALICE Collaboration acknowledges the following funding agencies for their support in building and running the ALICE detector:
A. I. Alikhanyan National Science Laboratory (Yerevan Physics Institute) Foundation (ANSL), State Committee of Science and World Federation of Scientists (WFS), Armenia;
Austrian Academy of Sciences and Nationalstiftung f\"{u}r Forschung, Technologie und Entwicklung, Austria;
Ministry of Communications and High Technologies, National Nuclear Research Center, Azerbaijan;
Conselho Nacional de Desenvolvimento Cient\'{\i}fico e Tecnol\'{o}gico (CNPq), Universidade Federal do Rio Grande do Sul (UFRGS), Financiadora de Estudos e Projetos (Finep) and Funda\c{c}\~{a}o de Amparo \`{a} Pesquisa do Estado de S\~{a}o Paulo (FAPESP), Brazil;
Ministry of Science \& Technology of China (MSTC), National Natural Science Foundation of China (NSFC) and Ministry of Education of China (MOEC) , China;
Ministry of Science, Education and Sport and Croatian Science Foundation, Croatia;
Ministry of Education, Youth and Sports of the Czech Republic, Czech Republic;
The Danish Council for Independent Research | Natural Sciences, the Carlsberg Foundation and Danish National Research Foundation (DNRF), Denmark;
Helsinki Institute of Physics (HIP), Finland;
Commissariat \`{a} l'Energie Atomique (CEA) and Institut National de Physique Nucl\'{e}aire et de Physique des Particules (IN2P3) and Centre National de la Recherche Scientifique (CNRS), France;
Bundesministerium f\"{u}r Bildung, Wissenschaft, Forschung und Technologie (BMBF) and GSI Helmholtzzentrum f\"{u}r Schwerionenforschung GmbH, Germany;
General Secretariat for Research and Technology, Ministry of Education, Research and Religions, Greece;
National Research, Development and Innovation Office, Hungary;
Department of Atomic Energy Government of India (DAE) and Council of Scientific and Industrial Research (CSIR), New Delhi, India;
Indonesian Institute of Science, Indonesia;
Centro Fermi - Museo Storico della Fisica e Centro Studi e Ricerche Enrico Fermi and Istituto Nazionale di Fisica Nucleare (INFN), Italy;
Institute for Innovative Science and Technology , Nagasaki Institute of Applied Science (IIST), Japan Society for the Promotion of Science (JSPS) KAKENHI and Japanese Ministry of Education, Culture, Sports, Science and Technology (MEXT), Japan;
Consejo Nacional de Ciencia (CONACYT) y Tecnolog\'{i}a, through Fondo de Cooperaci\'{o}n Internacional en Ciencia y Tecnolog\'{i}a (FONCICYT) and Direcci\'{o}n General de Asuntos del Personal Academico (DGAPA), Mexico;
Nederlandse Organisatie voor Wetenschappelijk Onderzoek (NWO), Netherlands;
The Research Council of Norway, Norway;
Commission on Science and Technology for Sustainable Development in the South (COMSATS), Pakistan;
Pontificia Universidad Cat\'{o}lica del Per\'{u}, Peru;
Ministry of Science and Higher Education and National Science Centre, Poland;
Korea Institute of Science and Technology Information and National Research Foundation of Korea (NRF), Republic of Korea;
Ministry of Education and Scientific Research, Institute of Atomic Physics and Romanian National Agency for Science, Technology and Innovation, Romania;
Joint Institute for Nuclear Research (JINR), Ministry of Education and Science of the Russian Federation and National Research Centre Kurchatov Institute, Russia;
Ministry of Education, Science, Research and Sport of the Slovak Republic, Slovakia;
National Research Foundation of South Africa, South Africa;
Centro de Aplicaciones Tecnol\'{o}gicas y Desarrollo Nuclear (CEADEN), Cubaenerg\'{\i}a, Cuba, Ministerio de Ciencia e Innovacion and Centro de Investigaciones Energ\'{e}ticas, Medioambientales y Tecnol\'{o}gicas (CIEMAT), Spain;
Swedish Research Council (VR) and Knut \& Alice Wallenberg Foundation (KAW), Sweden;
European Organization for Nuclear Research, Switzerland;
National Science and Technology Development Agency (NSDTA), Suranaree University of Technology (SUT) and Office of the Higher Education Commission under NRU project of Thailand, Thailand;
Turkish Atomic Energy Agency (TAEK), Turkey;
National Academy of  Sciences of Ukraine, Ukraine;
Science and Technology Facilities Council (STFC), United Kingdom;
National Science Foundation of the United States of America (NSF) and United States Department of Energy, Office of Nuclear Physics (DOE NP), United States of America.

%% file: Alice_Authorlist_2017-Jul-19.tex

\begingroup
\small
\begin{flushleft}
S.~Acharya\Irefn{org139}\And 
J.~Adam\Irefn{org99}\And 
D.~Adamov\'{a}\Irefn{org96}\And 
J.~Adolfsson\Irefn{org34}\And 
M.M.~Aggarwal\Irefn{org101}\And 
G.~Aglieri Rinella\Irefn{org35}\And 
M.~Agnello\Irefn{org31}\And 
N.~Agrawal\Irefn{org48}\And 
Z.~Ahammed\Irefn{org139}\And 
N.~Ahmad\Irefn{org17}
% ...
\And
\renewcommand\labelenumi{\textsuperscript{\theenumi}~}

\section*{Affiliation notes}
\renewcommand\theenumi{\roman{enumi}}
\begin{Authlist}
\item \Adef{org*}Deceased
\item \Adef{orgI}Dipartimento DET del Politecnico di Torino, Turin, Italy
\item \Adef{orgII}Georgia State University, Atlanta, Georgia, United States
\item \Adef{orgIII}M.V. Lomonosov Moscow State University, D.V. Skobeltsyn Institute of Nuclear, Physics, Moscow, Russia
\item \Adef{orgIV}Department of Applied Physics, Aligarh Muslim University, Aligarh, India
\item \Adef{orgV}Institute of Theoretical Physics, University of Wroclaw, Poland
\end{Authlist}

\section*{Collaboration Institutes}
\renewcommand\theenumi{\arabic{enumi}~}
\begin{Authlist}
\item \Idef{org1}A.I. Alikhanyan National Science Laboratory (Yerevan Physics Institute) Foundation, Yerevan, Armenia
\item \Idef{org2}Benem\'{e}rita Universidad Aut\'{o}noma de Puebla, Puebla, Mexico
\item \Idef{org3}Bogolyubov Institute for Theoretical Physics, Kiev, Ukraine
\item \Idef{org4}Bose Institute, Department of Physics  and Centre for Astroparticle Physics and Space Science (CAPSS), Kolkata, India
\item \Idef{org5}Budker Institute for Nuclear Physics, Novosibirsk, Russia
\item \Idef{org6}California Polytechnic State University, San Luis Obispo, California, United States
\item \Idef{org7}Central China Normal University, Wuhan, China
\item \Idef{org8}Centre de Calcul de l'IN2P3, Villeurbanne, Lyon, France
\item \Idef{org9}Centro de Aplicaciones Tecnol\'{o}gicas y Desarrollo Nuclear (CEADEN), Havana, Cuba
\item \Idef{org10}Centro de Investigaciones Energ\'{e}ticas Medioambientales y Tecnol\'{o}gicas (CIEMAT), Madrid, Spain
\item \Idef{org11}Centro de Investigaci\'{o}n y de Estudios Avanzados (CINVESTAV), Mexico City and M\'{e}rida, Mexico
\item \Idef{org12}Centro Fermi - Museo Storico della Fisica e Centro Studi e Ricerche ``Enrico Fermi', Rome, Italy
\item \Idef{org13}Chicago State University, Chicago, Illinois, United States
\item \Idef{org14}China Institute of Atomic Energy, Beijing, China
\item \Idef{org15}COMSATS Institute of Information Technology (CIIT), Islamabad, Pakistan
\item \Idef{org16}Departamento de F\'{\i}sica de Part\'{\i}culas and IGFAE, Universidad de Santiago de Compostela, Santiago de Compostela, Spain
\item \Idef{org17}Department of Physics, Aligarh Muslim University, Aligarh, India
\item \Idef{org18}Department of Physics, Ohio State University, Columbus, Ohio, United States
\item \Idef{org19}Department of Physics, Pusan National University, Pusan, Republic of Korea
\item \Idef{org20}Department of Physics, Sejong University, Seoul, Republic of Korea
\item \Idef{org21}Department of Physics, University of Oslo, Oslo, Norway
\item \Idef{org22}Department of Physics and Technology, University of Bergen, Bergen, Norway
\item \Idef{org23}Dipartimento di Fisica dell'Universit\`{a} 'La Sapienza' and Sezione INFN, Rome, Italy
\item \Idef{org24}Dipartimento di Fisica dell'Universit\`{a} and Sezione INFN, Cagliari, Italy
\item \Idef{org25}Dipartimento di Fisica dell'Universit\`{a} and Sezione INFN, Trieste, Italy
\item \Idef{org26}Dipartimento di Fisica dell'Universit\`{a} and Sezione INFN, Turin, Italy
\item \Idef{org27}Dipartimento di Fisica e Astronomia dell'Universit\`{a} and Sezione INFN, Bologna, Italy
\item \Idef{org28}Dipartimento di Fisica e Astronomia dell'Universit\`{a} and Sezione INFN, Catania, Italy
\item \Idef{org29}Dipartimento di Fisica e Astronomia dell'Universit\`{a} and Sezione INFN, Padova, Italy
\item \Idef{org30}Dipartimento di Fisica `E.R.~Caianiello' dell'Universit\`{a} and Gruppo Collegato INFN, Salerno, Italy
\item \Idef{org31}Dipartimento DISAT del Politecnico and Sezione INFN, Turin, Italy
\item \Idef{org32}Dipartimento di Scienze e Innovazione Tecnologica dell'Universit\`{a} del Piemonte Orientale and INFN Sezione di Torino, Alessandria, Italy
\item \Idef{org33}Dipartimento Interateneo di Fisica `M.~Merlin' and Sezione INFN, Bari, Italy
\item \Idef{org34}Division of Experimental High Energy Physics, University of Lund, Lund, Sweden
\item \Idef{org35}European Organization for Nuclear Research (CERN), Geneva, Switzerland
\item \Idef{org36}Excellence Cluster Universe, Technische Universit\"{a}t M\"{u}nchen, Munich, Germany
\item \Idef{org37}Faculty of Engineering, Bergen University College, Bergen, Norway
\item \Idef{org38}Faculty of Mathematics, Physics and Informatics, Comenius University, Bratislava, Slovakia
\item \Idef{org39}Faculty of Nuclear Sciences and Physical Engineering, Czech Technical University in Prague, Prague, Czech Republic
\item \Idef{org40}Faculty of Science, P.J.~\v{S}af\'{a}rik University, Ko\v{s}ice, Slovakia
\item \Idef{org41}Faculty of Technology, Buskerud and Vestfold University College, Tonsberg, Norway
\item \Idef{org42}Frankfurt Institute for Advanced Studies, Johann Wolfgang Goethe-Universit\"{a}t Frankfurt, Frankfurt, Germany
\item \Idef{org43}Gangneung-Wonju National University, Gangneung, Republic of Korea
\item \Idef{org44}Gauhati University, Department of Physics, Guwahati, India
\item \Idef{org45}Helmholtz-Institut f\"{u}r Strahlen- und Kernphysik, Rheinische Friedrich-Wilhelms-Universit\"{a}t Bonn, Bonn, Germany
\item \Idef{org46}Helsinki Institute of Physics (HIP), Helsinki, Finland
\item \Idef{org47}Hiroshima University, Hiroshima, Japan
\item \Idef{org48}Indian Institute of Technology Bombay (IIT), Mumbai, India
\item \Idef{org49}Indian Institute of Technology Indore, Indore, India
\item \Idef{org50}Indonesian Institute of Sciences, Jakarta, Indonesia
\item \Idef{org51}INFN, Laboratori Nazionali di Frascati, Frascati, Italy
\item \Idef{org52}INFN, Laboratori Nazionali di Legnaro, Legnaro, Italy
\item \Idef{org53}INFN, Sezione di Bari, Bari, Italy
\item \Idef{org54}INFN, Sezione di Bologna, Bologna, Italy
\item \Idef{org55}INFN, Sezione di Cagliari, Cagliari, Italy
\item \Idef{org56}INFN, Sezione di Catania, Catania, Italy
\item \Idef{org57}INFN, Sezione di Padova, Padova, Italy
\item \Idef{org58}INFN, Sezione di Roma, Rome, Italy
\item \Idef{org59}INFN, Sezione di Torino, Turin, Italy
\item \Idef{org60}INFN, Sezione di Trieste, Trieste, Italy
\item \Idef{org61}Inha University, Incheon, Republic of Korea
\item \Idef{org62}Institut de Physique Nucl\'eaire d'Orsay (IPNO), Universit\'e Paris-Sud, CNRS-IN2P3, Orsay, France
\item \Idef{org63}Institute for Nuclear Research, Academy of Sciences, Moscow, Russia
\item \Idef{org64}Institute for Subatomic Physics of Utrecht University, Utrecht, Netherlands
\item \Idef{org65}Institute for Theoretical and Experimental Physics, Moscow, Russia
\item \Idef{org66}Institute of Experimental Physics, Slovak Academy of Sciences, Ko\v{s}ice, Slovakia
\item \Idef{org67}Institute of Physics, Academy of Sciences of the Czech Republic, Prague, Czech Republic
\item \Idef{org68}Institute of Physics, Bhubaneswar, India
\item \Idef{org69}Institute of Space Science (ISS), Bucharest, Romania
\item \Idef{org70}Institut f\"{u}r Informatik, Johann Wolfgang Goethe-Universit\"{a}t Frankfurt, Frankfurt, Germany
\item \Idef{org71}Institut f\"{u}r Kernphysik, Johann Wolfgang Goethe-Universit\"{a}t Frankfurt, Frankfurt, Germany
\item \Idef{org72}Institut f\"{u}r Kernphysik, Westf\"{a}lische Wilhelms-Universit\"{a}t M\"{u}nster, M\"{u}nster, Germany
\item \Idef{org73}Instituto de Ciencias Nucleares, Universidad Nacional Aut\'{o}noma de M\'{e}xico, Mexico City, Mexico
\item \Idef{org74}Instituto de F\'{i}sica, Universidade Federal do Rio Grande do Sul (UFRGS), Porto Alegre, Brazil
\item \Idef{org75}Instituto de F\'{\i}sica, Universidad Nacional Aut\'{o}noma de M\'{e}xico, Mexico City, Mexico
\item \Idef{org76}IRFU, CEA, Universit\'{e} Paris-Saclay, Saclay, France
\item \Idef{org77}iThemba LABS, National Research Foundation, Somerset West, South Africa
\item \Idef{org78}Joint Institute for Nuclear Research (JINR), Dubna, Russia
\item \Idef{org79}Konkuk University, Seoul, Republic of Korea
\item \Idef{org80}Korea Institute of Science and Technology Information, Daejeon, Republic of Korea
\item \Idef{org81}KTO Karatay University, Konya, Turkey
\item \Idef{org82}Laboratoire de Physique Corpusculaire (LPC), Clermont Universit\'{e}, Universit\'{e} Blaise Pascal, CNRS--IN2P3, Clermont-Ferrand, France
\item \Idef{org83}Laboratoire de Physique Subatomique et de Cosmologie, Universit\'{e} Grenoble-Alpes, CNRS-IN2P3, Grenoble, France
\item \Idef{org84}Lawrence Berkeley National Laboratory, Berkeley, California, United States
\item \Idef{org85}Moscow Engineering Physics Institute, Moscow, Russia
\item \Idef{org86}Nagasaki Institute of Applied Science, Nagasaki, Japan
\item \Idef{org87}National and Kapodistrian University of Athens, Physics Department, Athens, Greece
\item \Idef{org88}National Centre for Nuclear Studies, Warsaw, Poland
\item \Idef{org89}National Institute for Physics and Nuclear Engineering, Bucharest, Romania
\item \Idef{org90}National Institute of Science Education and Research, HBNI, Jatni, India
\item \Idef{org91}National Nuclear Research Center, Baku, Azerbaijan
\item \Idef{org92}National Research Centre Kurchatov Institute, Moscow, Russia
\item \Idef{org93}Niels Bohr Institute, University of Copenhagen, Copenhagen, Denmark
\item \Idef{org94}Nikhef, Nationaal instituut voor subatomaire fysica, Amsterdam, Netherlands
\item \Idef{org95}Nuclear Physics Group, STFC Daresbury Laboratory, Daresbury, United Kingdom
\item \Idef{org96}Nuclear Physics Institute, Academy of Sciences of the Czech Republic, \v{R}e\v{z} u Prahy, Czech Republic
\item \Idef{org97}Oak Ridge National Laboratory, Oak Ridge, Tennessee, United States
\item \Idef{org98}Petersburg Nuclear Physics Institute, Gatchina, Russia
\item \Idef{org99}Physics Department, Creighton University, Omaha, Nebraska, United States
\item \Idef{org100}Physics department, Faculty of science, University of Zagreb, Zagreb, Croatia
\item \Idef{org101}Physics Department, Panjab University, Chandigarh, India
\item \Idef{org102}Physics Department, University of Cape Town, Cape Town, South Africa
\item \Idef{org103}Physics Department, University of Jammu, Jammu, India
\item \Idef{org104}Physics Department, University of Rajasthan, Jaipur, India
\item \Idef{org105}Physikalisches Institut, Eberhard Karls Universit\"{a}t T\"{u}bingen, T\"{u}bingen, Germany
\item \Idef{org106}Physikalisches Institut, Ruprecht-Karls-Universit\"{a}t Heidelberg, Heidelberg, Germany
\item \Idef{org107}Physik Department, Technische Universit\"{a}t M\"{u}nchen, Munich, Germany
\item \Idef{org108}Purdue University, West Lafayette, Indiana, United States
\item \Idef{org109}Research Division and ExtreMe Matter Institute EMMI, GSI Helmholtzzentrum f\"ur Schwerionenforschung GmbH, Darmstadt, Germany
\item \Idef{org110}Rudjer Bo\v{s}kovi\'{c} Institute, Zagreb, Croatia
\item \Idef{org111}Russian Federal Nuclear Center (VNIIEF), Sarov, Russia
\item \Idef{org112}Saha Institute of Nuclear Physics, Kolkata, India
\item \Idef{org113}School of Physics and Astronomy, University of Birmingham, Birmingham, United Kingdom
\item \Idef{org114}Secci\'{o}n F\'{\i}sica, Departamento de Ciencias, Pontificia Universidad Cat\'{o}lica del Per\'{u}, Lima, Peru
\item \Idef{org115}SSC IHEP of NRC Kurchatov institute, Protvino, Russia
\item \Idef{org116}Stefan Meyer Institut f\"{u}r Subatomare Physik (SMI), Vienna, Austria
\item \Idef{org117}SUBATECH, IMT Atlantique, Universit\'{e} de Nantes, CNRS-IN2P3, Nantes, France
\item \Idef{org118}Suranaree University of Technology, Nakhon Ratchasima, Thailand
\item \Idef{org119}Technical University of Ko\v{s}ice, Ko\v{s}ice, Slovakia
\item \Idef{org120}Technical University of Split FESB, Split, Croatia
\item \Idef{org121}The Henryk Niewodniczanski Institute of Nuclear Physics, Polish Academy of Sciences, Cracow, Poland
\item \Idef{org122}The University of Texas at Austin, Physics Department, Austin, Texas, United States
\item \Idef{org123}Universidad Aut\'{o}noma de Sinaloa, Culiac\'{a}n, Mexico
\item \Idef{org124}Universidade de S\~{a}o Paulo (USP), S\~{a}o Paulo, Brazil
\item \Idef{org125}Universidade Estadual de Campinas (UNICAMP), Campinas, Brazil
\item \Idef{org126}Universidade Federal do ABC, Santo Andre, Brazil
\item \Idef{org127}University of Houston, Houston, Texas, United States
\item \Idef{org128}University of Jyv\"{a}skyl\"{a}, Jyv\"{a}skyl\"{a}, Finland
\item \Idef{org129}University of Liverpool, Liverpool, United Kingdom
\item \Idef{org130}University of Tennessee, Knoxville, Tennessee, United States
\item \Idef{org131}University of the Witwatersrand, Johannesburg, South Africa
\item \Idef{org132}University of Tokyo, Tokyo, Japan
\item \Idef{org133}University of Tsukuba, Tsukuba, Japan
\item \Idef{org134}Universit\'{e} de Lyon, Universit\'{e} Lyon 1, CNRS/IN2P3, IPN-Lyon, Villeurbanne, Lyon, France
\item \Idef{org135}Universit\'{e} de Strasbourg, CNRS, IPHC UMR 7178, F-67000 Strasbourg, France, Strasbourg, France
\item \Idef{org136}Universit\`{a} degli Studi di Pavia, Pavia, Italy
\item \Idef{org137}Universit\`{a} di Brescia, Brescia, Italy
\item \Idef{org138}V.~Fock Institute for Physics, St. Petersburg State University, St. Petersburg, Russia
\item \Idef{org139}Variable Energy Cyclotron Centre, Kolkata, India
\item \Idef{org140}Warsaw University of Technology, Warsaw, Poland
\item \Idef{org141}Wayne State University, Detroit, Michigan, United States
\item \Idef{org142}Wigner Research Centre for Physics, Hungarian Academy of Sciences, Budapest, Hungary
\item \Idef{org143}Yale University, New Haven, Connecticut, United States
\item \Idef{org144}Yonsei University, Seoul, Republic of Korea
\item \Idef{org145}Zentrum f\"{u}r Technologietransfer und Telekommunikation (ZTT), Fachhochschule Worms, Worms, Germany
\end{Authlist}
\endgroup